\documentclass[prd,showpacs,nofootinbib]{revtex4}

\usepackage{amsmath}
\usepackage{amsfonts}
\usepackage{bm}
\usepackage{graphics}

\newcommand{\be}{\begin{equation}}
\newcommand{\ee}{\end{equation}}
\newcommand{\ba}{\begin{eqnarray}}
\newcommand{\ea}{\end{eqnarray}}
\newcommand{\sH}{{\text{\tiny $\phi H$}}}
\newcommand{\sV}{{\text{\tiny $\phi V$}}}
\newcommand{\tH}{{\text{\tiny $TH$}}}
\newcommand{\tV}{{\text{\tiny $TV$}}}
\renewcommand{\H}{{\text{\tiny $H$}}}
\newcommand{\V}{{\text{\tiny $V$}}}

\begin{document}

\title{Slow-roll parameters in braneworld cosmologies}
\author{Gianluca Calcagni}
\email{calcagni@fis.unipr.it}
\affiliation{Dipartimento di Fisica ``M. Melloni,'' Universit\`{a} di Parma, Parco Area delle Scienze 7/A, I-43100 Parma,
 Italy}
\date{February 11, 2004}
\begin{abstract}
We present a general slow-roll formalism within braneworld-motivated cosmologies with non-standard effective Friedmann equations. Full towers of parameters involving either the inflaton potential or the Hubble parameter are constructed and the dynamics of non-tachyonic and tachyonic fields are considered in detail; exact cosmological solutions and the inflationary attractor condition are provided. We compare scalar-driven and tachyon-driven accelerating eras through 
slow-roll correspondence and the observational imprint on early-universe structures.
\end{abstract}

\preprint{UPRF-2004-01}

\maketitle


\section{Introduction}
 
Motivated by recent developments of string, superstring and M theories, several models for a multidimensional target spacetime have been proposed \cite{HW1,HW2,ADD98,AADD,LOSW,RSa,RSb}. Soon they found important applications in cosmology by revitalizing the idea that the visible universe is a (3+1)-dimensional variety (a 3-brane) embedded in a bulk with some extra dimensions, either non-compact or compactified \cite{aka82,RuS,vis85}. Typically, the background metric on the brane is the Friedmann-Robertson-Walker (FRW) metric and the Einstein equations are modified in accordance with the gravity model permeating the whole spacetime. This in turn produces the basic FRW equations for the cosmological evolution; for a comprehensive review, see \cite{rub01,mar03}. In particular, the Randall-Sundrum scenarios \cite{RSa,RSb} and their Gauss-Bonnet generalization \cite{KKL1,KKL2} (see also \cite{LN,GrP} and references therein) seem promising because of their high-energy features, while preserving the standard cosmology in the low-energy limit.
             
In order to reconcile cosmological and astrophysical observations with the standard big bang scenario, it is sufficient (but not necessary) to invoke the inflationary paradigm; according to it, the early Universe experienced a phase of accelerated expansion driven by an effective cosmological constant. This mechanism is triggered by the dynamics of a scalar field rolling down its potential and may also provide an explanation for the present phase of acceleration; in the most famous version of inflation, the rolling is slow enough to justify the adoption of the slow-roll (SR) formalism. For a review of primordial 4D inflation and the SR approach, see \cite{lid97}. Recently, due to many progresses made in understanding the vacuum structure of string theory (in particular, see \cite{sen1,sen2,sen3,sen4,HKM,GS,KMM1,KMM2,sen5,sen6,sen7,sen8}), the eventuality that the scalar field is tachyonic has been explored.

In this paper we develop a general SR formalism starting from a FRW equation which is polynomial in the brane energy density, $H^2 \propto \rho^q$; by this way we obtain a Hamilton-Jacobi and SR formulation of the cosmological evolution which is valid for many known braneworld theories either in a particular energy limit or time interval. We will restrict ourselves to a brane filled (1) with a scalar or (2) with a tachyon field with an effective Dirac-Born-Infeld action provided by the low coupling limit of non-perturbative string theory. The advantages of this approach are several. It provides a concise and versatile formalism to explore different cosmological models and determine their main features, such as exact classes of solutions, the inflationary attractor and the inflationary imprint on the structure formation of the early Universe. 

A general example of such a braneworld-generated cosmological equation is given in \cite{ChF}. This model has been used to describe the post-inflationary evolution and in this case has been dubbed ``Cardassian cosmology'' \cite{car1,car2,car3,car4,car5,car6,car7,car8,car9,car10,car11,car12}. Here, we will take a rather different perspective and ask how a period of non-standard expansion can modify the usual early-Universe picture.
 
The plan of this work is the following: in Sec. \ref{background} we motivate the ``patch cosmology'' approach and set the background equations, both for a scalar and a tachyon field. Section \ref{slowroll} is devoted to the slow-roll formalism and its consequences in the scalar and tachyon scenarios. The inflationary attractor is then considered in Sec. \ref{attractor}, while in Sec. \ref{exact} we construct some exact cosmological solutions; these sections are presented in a self-contained manner. In Sec. \ref{pert} we calculate next-to-lowest order scalar spectral amplitudes and indices and compare 4D and Randall-Sundrum models through consistency equations. The intertwine between scalar-driven and tachyon-driven cosmologies is stressed in Sec. \ref{vs} and in the Appendix we use the slow-roll correspondence between tachyonic and non-tachyonic dynamics to estimate the non-Gaussianity produced during a four-dimensional tachyon-driven inflationary period. Conclusions are in Sec. \ref{concl}.


\section{Background cosmology} \label{background}


\subsection{Gauss-Bonnet braneworld and energy patches}

\begin{figure}[ht]
\includegraphics{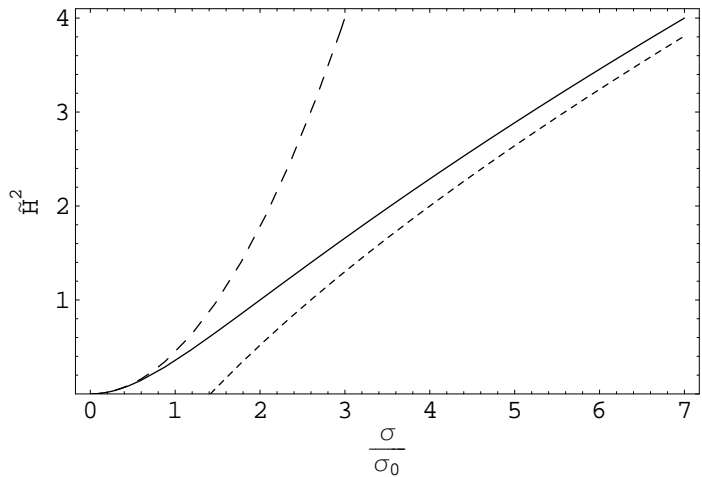}
\caption{\label{fig1} The Hubble parameter as a function of the energy density in the Gauss-Bonnet scenario and its energy approximations. The solid line is the full Gauss-Bonnet cosmology [Eq. (\ref{gabo})], the lower dashed line is the high-energy Gauss-Bonnet regime [Eq. (\ref{fe1})] and the upper dashed line is the full Randall-Sundrum regime.}
\end{figure}

Let us start with the five-dimensional bulk action for the Gauss-Bonnet braneworld:
\be
S =\frac{1}{2\kappa_5^2} \int_{\cal M} d^5x\sqrt{-g_5}\left[R-2\Lambda_5+\alpha\left(R^2-4R_{\mu\nu}R^{\mu\nu}+R_{\mu\nu\rho\sigma}R^{\mu\nu\rho\sigma}\right)\right]+S_\partial+S_\text{\tiny matter}\,.
\ee
Here $\kappa_5^2=8\pi/m_5^3$ is the five-dimensional gravitational coupling, $g_5$ is the determinant of the 5D metric, $\Lambda_5<0$ is the bulk cosmological constant and $\alpha=1/(8g_s)>0$ is the Gauss-Bonnet coupling, where $g_s$ is the string energy scale. The action includes a pure geometrical boundary term $S_\partial$ and the matter contribution which is confined on the brane. The gravitational part of the action is a natural generalization in five dimensions of the Einstein-Hilbert action (see \cite{CD} for a general discussion) and, from a fundamental physics point of view, it comes from zeroth + leading order stringy corrections to gravity. Assuming a perfect fluid matter and a $\bm{Z}_2$ symmetry across the brane, the effective Friedmann equation on the brane is \cite{CD,dav03,GW}
\begin{subequations} \label{gabo}
\be 
H^2=\frac{c_+ + c_- -2}{8\alpha}\,,
\ee
where $H$ is the Hubble parameter and, defining $\sqrt{\alpha/2}\,\kappa_5^2 \equiv \sigma_0^{-1}$,
\be
c_\pm = \left\{\left[\left(1+\frac{4}{3}\alpha\Lambda_5\right)^{3/2}+\left(\frac{\sigma}{\sigma_0}\right)^2\right]^{1/2} \pm \frac{\sigma}{\sigma_0}\right\}^{2/3}\,;
\ee
\end{subequations}
$\sigma$ is the matter energy density which we will assume to be decomposed into a matter contribution plus the brane tension $\lambda$: $\sigma=\rho+\lambda$. Expanding Eq. (\ref{gabo}) to quadratic order in $\sigma$ one recovers the Friedmann equation of the Randall-Sundrum type II scenario with vanishing 4D cosmological constant \cite{CGS,CGKT,BDL,BDEL,FTW}, provided
\[\kappa_5^4=\frac{6\kappa_4^2}{\lambda}\left(1+\frac{4}{3}\Lambda_5\alpha\right)\,,\qquad \lambda=\frac{3}{2\alpha\kappa_4^2}\left[1-\left(\frac{\lambda \kappa_5^4}{6 \kappa_4^2}\right)^{1/2}\right]\,,\]
where $\kappa_4^2=8\pi/m_4^2$ and $m_4$ is the four-dimensional Planck mass. We can now recognize three main energy regimes resulting in particular limits of the Friedmann equation:
\begin{enumerate}
	\item $\sigma/\sigma_0 \gg 1\,:$ In this ``pure Gauss-Bonnet'' high-energy regime, we have a non-standard cosmology
	      \be \label{fe1}
	      H^2= \left(\frac{\kappa_5^2}{16\alpha}\right)^{2/3}\sigma^{2/3}\,.
	      \ee
	\item $\lambda/\sigma_0 \ll \sigma/\sigma_0 \ll 1\,:$ When the energy density is far below the 5D or string scale but
	      $\rho \gg \lambda$, we have a Hubble parameter
	      \be \label{fe2}
	      H^2 = \frac{\kappa_4^2}{6 \lambda} \rho^2\,.
	      \ee
	\item $\rho/\sigma_0 \ll \sigma/\sigma_0 \ll 1\,:$ The standard four-dimensional scenario is recovered when the brane
	      grows stiff with respect to its matter content, $\rho \ll \lambda$:
	      \be \label{fe3}
        H^2 = \frac{\kappa_4^2}{3} \rho \,.
        \ee
\end{enumerate}
The Friedmann equation (\ref{gabo}) and its energy approximations are plotted in Fig. \ref{fig1}.

Equations (\ref{fe1})--(\ref{fe3}) are considerably simpler than the full Gauss-Bonnet equation (\ref{gabo}) and in many practical cases one of the three energy regimes is assumed. Therefore, it can be useful to study a cosmological patch, that is a region of time and energy in which $H^2 \propto \rho^q$ holds for some constant $q$, and apply the resulting general equations to any case of interest, let it be a particular limit of either the Gauss-Bonnet braneworld or another scenario where the visible Universe is confined on a four-dimensional variety embedded in a higher-dimensional spacetime. The parameter $q$, which describes the effective degrees of freedom from gravity, could live in a non-standard range of values because of the introduction of non-perturbative stringy effects or, just to mention some possibilities, for the presence of a complicated geometrical framework with either compact and non-compact extra dimensions, multiple and/or folding branes configurations, and so on. Even if this were not the case, a patch formulation of the cosmological problem would provide a compact notation for many situations. Of course, this approach will be valid far from transitions between patches; in fact, the Hubble parameter will be a more or less complicated function of the energy density, say Eq. (\ref{gabo}), with smooth transitions from an energy (and SR) regime to the other.

\begin{figure}[ht]
\includegraphics{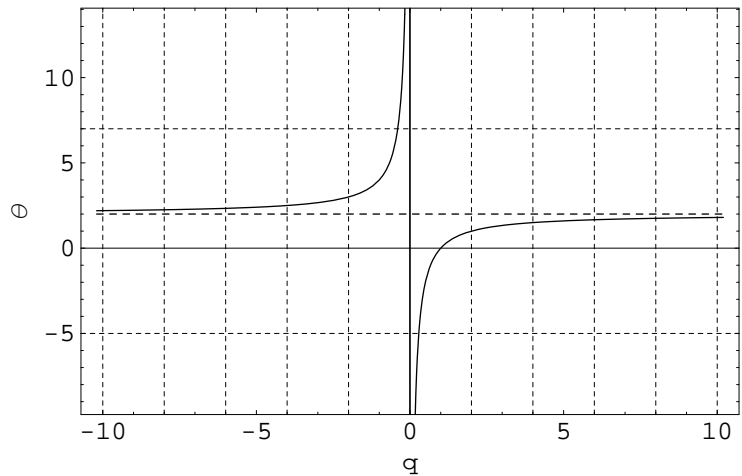}
\caption{\label{fig2} The parameter $\theta (q)$. The three cosmologies described in the text are: high-energy Gauss-Bonnet [$\theta (2/3)=-1$], standard four-dimensional [$\theta (1) =0$] and high-energy Randall-Sundrum [$\theta (2)=1$].}
\end{figure}

In the rest of this section, we investigate the properties of a single flat energy patch with general effective Friedmann equation
\be \label{FRW}
H^2=\beta_q^2 \rho^q\,,
\ee
where $q$ is constant and $\beta_q>0$ is some factor with energy dimension $[\beta_q]= E^{1-2q}$. We neglect any contribution from the Weyl tensor and assume, without further analysis, there is some confinement mechanism for a perfect fluid with equation of state $p=w\rho$ and continuity equation
\be \label{conti}
\dot{\rho}+3H (\rho+p) = 0\,.
\ee
This is equivalent to the local covariant conservation of the energy-momentum tensor. Differentiating Eq. (\ref{FRW}) with respect to time and using Eq. (\ref{conti}), one gets
\ba
\dot{\widetilde{H}} &=& -\frac{3}{2} q \beta_q \widetilde{H}^\theta (\rho+p) \label{dotH1}\\
                    &=& -\frac{3}{2} q \beta_q \widetilde{H}^2 (1+w)\,, \label{dotH0}
\ea
where $\widetilde{H}\equiv H/\beta_q$ and the exponent
\be
\theta \equiv 2\left(1-\frac{1}{q}\right)\,,
\ee
is shown in Fig. \ref{fig2}.

If one imposes the dominant energy condition, $\rho \geq |p|$, Eq. (\ref{dotH0}) states that the Hubble length\footnote{The Hubble length is the proper distance from the observer of an object the observer sees moving with the cosmological expansion at the velocity of light.} $R_H=H^{-1}$ is monotonic during its time evolution, increasing if $q>0$ (the lower branch with $\theta <2$) and decreasing if $q<0$ ($\theta>2$). On the contrary, the particle horizon $R_c(t)=a(t)\int_{t_0}^t dt'/a(t')$, which defines the causally connected region centered in the observer, is always increasing in an expanding universe, $\dot{R_c}=1+HR_c>0$, and its comoving counterpart is always increasing also, $(R_c/a)^{^{\bm .}}>0$. We will denote with a subscript $0$ any quantity evaluated at the reference initial time $t_0$. For a constant index $w$, 
\be \label{conw}
\rho=\rho_0\left(\frac{a}{a_0}\right)^{-3(1+w)}\,,
\ee
while the scale factor is
\be
a(t) = a_0 \left[1+\epsilon_\H H_0(t-t_0)\right]^{1/\epsilon_\H}\,,
\ee
where 
\be\label{epsilon}
\epsilon_\H \equiv -\frac{\dot{H}}{H^2}= \frac{3q(1+w)}{2}\,.
\ee 
Thus the upper branch solution satisfying the dominant energy condition represents a super-inflationary\footnote{With standard FRW equations, $q=1$, super-inflationary models are those with $w<-1$ \cite{LM1}.} ($\dot{H}>0$) expanding ($H>0$) universe in a ``pre-big-bang'' era with time running from $t_0$ to eventually $\bar{t}=t_0-1/(\epsilon_\H H_0)>t_0$, where one encounters a singularity with infinite scale factor and vanishing energy density. Since any patch should be regarded as a model with a limited time interval of validity, its long-range evolution is a true problem only for well-established regimes. Anyway, we will consider only positive $q$, which is the case of all the known realistic cosmologies, and come back to this issue in Sec. \ref{concl}.

Whenever cosmological equations can be applied both in the scalar and tachyon case, the inflaton field will be generically indicated as $\psi$. Table \ref{table1} summarizes the three main cosmological regimes. For completeness, we have also shown the de Sitter (dS) solution with constant Hubble parameter, when $H=\beta_0=\text{constant}$, $\ddot{a}/a=\beta_0^2>0$ and inflation is driven by a cosmological constant with equation of state $w=-1$. Also, this case can be obtained via the formal limit $q \rightarrow 0$. The de Sitter regime is the idealization of the extreme slow-roll (ESR) approximation, $\dot{\psi} \approx 0$. In this regime, the kinetic term of the action is strongly subdominant with respect to the potential itself and $\widetilde{H} \approx V^{q/2}$.
\begin{table}[ht]
\caption{\label{table1}The energy regimes described in the text. The de Sitter case can be seen as the asymptotic cosmology with $q \rightarrow 0$. Here, $\epsilon_\text{max}=w_\text{max}+1$ is the maximum value for the index of the equation of state allowing inflation. Note that, in the Gauss-Bonnet regime, $\epsilon=\epsilon_\H$.}
\begin{ruledtabular}
\begin{tabular}{cccccc}
Regime &   $q$   &   $\theta$  &          $\beta_q^2$           & $\epsilon_\text{\tiny max}$ \\ \hline
dS     &    0    &   $\infty$  &             $H^2$              &                 $0$         \\
GB     &  $2/3$  &      $-1$   & $(\kappa_5^2/16\alpha)^{2/3}$  &                 $1$         \\
4D     &    1    &       $0$   &         $\kappa_4^2/3$         &             $2/3$           \\
RS2    &    2    &       $1$   &      $\kappa_4^2/6\lambda$     &             $1/3$           \\
\end{tabular}\end{ruledtabular}
\end{table}


\subsection{Patch cosmology with a scalar field}

In the following we will consider an inflationary four-dimensional flat universe filled with an homogeneous scalar field with energy density and pressure
\be \label{rhop}
\rho = \frac{\dot{\phi}^2}{2} + V(\phi) = p+2V(\phi)\,,
\ee
and effective equation of motion
\be
\ddot{\phi}+3H \dot{\phi}+ V'=0\,,\label{eom0}
\ee
where dots stand for synchronous-time derivatives and a prime denotes $\phi$ derivative. From Eq. (\ref{dotH1}),
\be
\dot{\widetilde{H}} = -\frac{3}{2} q \beta_q \widetilde{H}^{\theta} \dot{\phi}^2\,;\label{dotH}
\ee
equivalently, we can regard $H$ as a function of $\phi$:
\be \label{hj2}
\frac{\widetilde{H}'}{\widetilde{H}^\theta} = -\frac{3}{2} q \beta_q \dot{\phi}\,,
\ee
where the last passage is possible if $\phi$ varies monotonically with time.
Equations (\ref{FRW}), (\ref{conti}) and (\ref{rhop}) then give
\be \label{hj3}
\frac{2}{(3q\beta_q)^2}\frac{\widetilde{H}'^2}{\widetilde{H}^{2\theta}}-\widetilde{H}^{2-\theta}+V=0\,.
\ee
Equations (\ref{FRW}), (\ref{hj2}) and (\ref{hj3}) are the Hamilton-Jacobi equations; they are in agreement with the equations found in the low energy limit (4D) \cite{Mus90,SB}, in the Randall-Sundrum high-energy limit (RS2) \cite{HaL1} and in the Gauss-Bonnet high-energy limit (GB) \cite{MW}.


\subsection{Patch cosmology with a tachyon field}
 
The deep interplay between small-scale non-perturbative string theory (especially in the effective Born-Infeld action formulation) and large-scale braneworld scenarios has raised the interest in a tachyon field as an inflationary mechanism \cite{MPP,ale02}. Subsequently, the problem has been studied in a more cosmological fashion \cite{gib02,FT,muk02,fei02,pad02,SW,CGJP,FKS,KL,CFM,GPCZ,HN,SCQ,HL,ben02,LLH,sam03,WAS,BBS,gib03,CF,AF,BSS,SV,GKMP,PS}. Throughout this paper, a ``tachyon'' is by definition any scalar field $T$ with effective action $S=\int d^4x\, {\cal L}$ and Lagrangian \cite{gib02,sen4,gar00,ber00,klu00,GHY}
\be \label{ta}
{\cal L}= -V(T) \sqrt{-\det[g_{\mu\nu}-f(T) \partial_\mu T \partial_\nu T]}\,.
\ee
Here, $g_{\mu\nu}$ is the induced four-dimensional FRW metric on the brane, $T$ is the real tachyon field with dimension $[T]=E^{-1}$, $f$ is a function of $T$ and $V>0$ is the potential, which is exact to all orders in Regge slope but at the tree level in $g_s$. In the following, without loss of generality we will assume that $f(T)=1$ and $T=T(t)$ is homogeneous and monotonic, say, $\dot{T}>0$. In the case of a $D-\bar{D}$ system, the field $T$ is complex due to the Chan-Paton structure, but many of the following arguments will hold in this case too. Often, the tensor $G_{\mu\nu}=g_{\mu\nu}- \partial_\mu T \partial_\nu T$ is called the tachyon metric. Moreover, we will leave the exact form of the potential unspecified, except in Sec. \ref{exact}; if $V$ were constant, the model would correspond to a brane filled with Chaplygin gas, $p = -V^2/\rho$, which at late times behaves as an effective cosmological constant (e.g., \cite{jac00,KMP}). Otherwise, in general the potential will have a maximum at $T_0=0$ and a local minimum $V(T_*)=0$  either at $T_*$ finite or at infinity. In the latter case, there are no oscillations and a reheating mechanism appears difficult \cite{KL,CFM}.

The tachyon energy density and pressure read, respectively,
\ba
\rho &=& \frac{V(T)}{\sqrt{1-\dot{T}^2}}\,,\label{Trho}\\
p &=& -V(T) \sqrt{1-\dot{T}^2}=-\frac{V(T)^2}{\rho}\,.
\ea
Note that when $\dot{T}^2 \rightarrow 1$, the tachyon behaves as a pressureless gas. The continuity equation (\ref{conti}) gives the equation of motion
\be \label{Teom}
\frac{\ddot{T}}{1-\dot{T}^2}+3H\dot{T}+(\ln V)'=0\,,
\ee
where $V$ is differentiated with respect to $T$. Equation (\ref{dotH0}) then gives
\be \label{Thj2}
\frac{\widetilde{H}'}{\widetilde{H}^2} = -\frac{3}{2} q \beta_q \dot{T}\,.
\ee
By this equation and Eqs. (\ref{FRW}) and (\ref{Trho}), we have
\be \label{Thj3}
\frac{4}{(3q\beta_q)^2}\frac{\widetilde{H}'^2}{\widetilde{H}^{2\theta}}-\widetilde{H}^{2(2-\theta)}+V^2=0\,.
\ee
Equations (\ref{FRW}), (\ref{Thj2}) and (\ref{Thj3}) are the Hamilton-Jacobi equations for the tachyon; they agree with \cite{MW,FT}.


\section{Slow-roll formalism in braneworld cosmologies} \label{slowroll}

According to the inflationary idea, an era of accelerated expansion is driven by a scalar field slowly ``rolling'' down its potential into a local minimum. The use of the slow-roll formalism \cite{ST,LL,KV,LPB} simplifies the study of many consequences of inflation; however, it can also be considered as an effective notation for some recurrent dimensionless combinations of cosmological quantities, without imposing any condition on their magnitude. We will keep calling these parameters ``slow-roll'' in this case, too. The most commonly used SR towers rely upon two different quantities, the geometrical Hubble parameter $H$ and the dynamical inflaton potential $V$. We will name these towers H-SR and V-SR, respectively, and explore some of their properties in the general cosmology (\ref{FRW}). Other SR towers can be constructed for particular cosmological scenarios or analyses \cite{LN,STG,kin02,HaL2,RL}. We will also consider what happens in the case of a tachyonic field.


\subsection{H-SR parameters for a scalar field}

The H-SR tower is defined as
\begin{subequations} \label{hsr}
\ba
\epsilon_{\text{\tiny $\phi H$},0} &\equiv& \epsilon_\H = -\frac{d \ln H}{d \ln a}\,,\\
\epsilon_{\text{\tiny $\phi H$},n} &\equiv& \prod_{i=1}^n \left\{-\frac{d \ln \left[(H'H^{-\theta})^{(i-1)}\right]}{d \ln a}\right\}^{1/n}\,, \qquad n \geq 1\,,
\ea
\end{subequations}
where $(n)$ is the $n$-th $\phi$ derivative. For a scalar field, the first three parameters, which are those appearing in all the main expressions for cosmological observables, are
\ba
\epsilon_\sH &\equiv& \epsilon_{\text{\tiny $\phi H$},0} =3q \frac{\dot{\phi}^2/2}{V+\dot{\phi}^2/2}\,, \label{phepsilon}\\
\eta_\sH     &\equiv& \epsilon_{\text{\tiny $\phi H$},1}   = -\frac{d \ln \dot{\phi}}{d \ln a}=-\frac{\ddot{\phi}}{H\dot{\phi}} \label{eta}\,,\\
\xi^2_\sH    &\equiv& \epsilon_{\text{\tiny $\phi H$},2}^2 =  \frac{1}{H^2} \left(\frac{\ddot{\phi}}{\dot{\phi}}\right)^. = \frac{\dddot{\phi}}{H^2\dot{\phi}}- \eta_\sH^2\,.\label{xi}
\ea
For $q>0$, the Hubble length is always non-decreasing; therefore, $\epsilon_\H = \dot{R}_\H \geq 0$. Moreover, we have a precise definition for the beginning of the inflationary era, since $\ddot{a}/a = H^2 (1-\epsilon_\H)>0$ iff $\epsilon_\H(t) < 1$; the end of inflation is set by $\epsilon_\H(t_*)=1$. The condition $\epsilon_\H \ll 1$ (ESR regime) permits to neglect the first term in the left hand side of Eq. (\ref{hj3}), that can be recast as 
\be \label{hjalt}
V= \left(1-\frac{1}{3q}\epsilon_\H\right)\widetilde{H}^{2/q}\,,
\ee
while $|\eta_\H| \ll 1$ is equivalent to assume the attractor solution $\dot{\phi} \approx -V'/3H$ from Eq. (\ref{eom0}). Consequently, all the dynamical information is encoded in the SR parameters. 
Equation (\ref{dotH}) can be rewritten in terms of $\epsilon_\H$, giving
\be \label{useful1}
\dot{\phi}^2=\frac{2}{3q}\widetilde{H}^{2/q}\epsilon_\H\,.
\ee
From Eq. (\ref{epsilon}), it follows that
\be \label{w}
w=\frac{2}{3q}\,\epsilon_\H-1\,;
\ee
thus the scalar field behaves almost like an effective cosmological constant in the SR approximation, $w \gtrsim -1$. The condition for inflation is then $w<-1+2/3q$.

Noting that $\ddot{H}=2H\dot{H}[(-1+1/q)\,\epsilon_\H-\eta_\H]$, we have
\ba
\dot{\epsilon}_\sH &=& 2H\epsilon_\sH \left(\frac{1}{q}\,\epsilon_\sH-\eta_\sH\right)\,,\label{epsih'}\\
\dot{\eta}_\sH     &=&  H\left(\epsilon_\sH\eta_\sH-\xi_\sH^2\right)\,.\label{etah'}
\ea
Differentiation with respect to the scalar field yields $\epsilon_{\text{\tiny $\phi H$},n}'=\dot{\epsilon}_{\text{\tiny $\phi H$},n}/\dot{\phi}\,$; by Eq. (\ref{useful1}), the resulting prefactor $H/\dot{\phi}$ can be expressed as
\be \label{usefull}
\frac{H}{\dot{\phi}}=+\left(\frac{3q\beta_q^2}{2} \frac{\widetilde{H}^\theta}{\epsilon_\H}\right)^{1/2}\,,
\ee
where the plus sign has been chosen in order to have a slow rolling down the potential with $\dot{\phi}>0$. This is always possible by a redefinition $\phi \rightarrow -\phi$.

A final comment is in order: when defined, a SR tower is dynamical (i.e., does say something about the dynamics of the Hamilton-Jacobi equations) either when constraints on the form and magnitude of the SR parameters are applied, or when distinct SR definitions are related through the Hamilton-Jacobi equations themselves. For example, the H-SR tower relies on the parameter $\epsilon_\H$, which is its fundamental ground; as far as one does not assume any specific link between the Hubble parameter (and its derivatives) and the fields living on the brane, it is clear there will be no knowledge about the evolution of the system. However, when rewriting these H-parameters in terms of $\dot{\phi}$ through the second Hamilton-Jacobi equation, these parameters become dynamical. This ambiguity may lead to confusion in some situations; an interesting discussion on related issues can be found in \cite{lid03}.


\subsection{V-SR parameters for a scalar field}

The H-SR hierarchy is an elegant instrument of analysis coming from the Hamilton-Jacobi formulation of the equations of motion. However, in many cases investigation starts from the inflaton potential $V$ and not from the Hubble parameter, whose shape must be determined by the Hamilton-Jacobi equations which are not always readily solvable. So, it is convenient to define another SR tower and try to relate it to the original one, namely, 
\begin{subequations} \label{vsr}
\ba
\epsilon_{\text{\tiny $\phi V$},0} &\equiv& \frac{q}{6\beta_q^2}\frac{V'^2}{V^{1+q}}\,,\\
\epsilon_{\text{\tiny $\phi V$},n} &\equiv& \frac{1}{3\beta_q^2}\left[\frac{V^{(n+1)}(V')^{n-1}}{V^{nq}}\right]^{1/n}\,, \qquad n \geq 1\,,
\ea
\end{subequations}
where again we have introduced by hand the first parameter. Therefore,
\ba
\epsilon_\sV &\equiv& \epsilon_{\text{\tiny $\phi V$},0}\,,\\
\eta_\sV     &\equiv& \epsilon_{\text{\tiny $\phi V$},1} =\frac{1}{3\beta_q^2}\frac{V''}{V^q}\,,\\
\xi_\sV^2    &\equiv& \epsilon_{\text{\tiny $\phi V$},2}^2=\frac{1}{(3\beta_q^2)^2}\frac{V'''V'}{V^{2q}}\,,
\ea
and their derivatives with respect to the scalar field are
\ba
\epsilon_\sV' &=& -q \frac{V'}{V} \left[\left(1+\frac{1}{q}\right)\epsilon_\sV-\eta_\sV\right]\,,\label{epsiv'}\\
\eta_\sV'     &=& -\frac{q}{2\epsilon_\sV} \frac{V'}{V} \left[2\epsilon_\sV\eta_\sV-\xi_\sV^2\right]\,,
\ea
where
\be \label{usefulV}
-\frac{V'}{V} = \left(\frac{6\beta_q^2}{q} \epsilon_\sV V^{q-1}\right)^{1/2}\,.
\ee
The conditions $\epsilon_\V \ll 1$ and $|\eta_\V| \ll 1$ are necessary to drop the kinetic term in Eq. (\ref{FRW}) and the acceleration term in Eq. (\ref{eom0}) but they are not sufficient. In general, this SR formalism requires a further assumption, namely, $\dot{\phi} \approx -V'/3H$, which is easy enough to be satisfied. This determines the minus sign in Eq. (\ref{usefulV}), provided $\dot{\phi}>0$.


\subsection{H-SR parameters for a tachyon}

In the tachyonic case, the first H-SR parameters are [see Eq. (\ref{epsilon})]
\ba
\epsilon_\H &=& \frac{3q}{2} \dot{T}^2\,, \label{Tepsilon}\\
\eta_\H&=&-\frac{\ddot{T}}{H\dot{T}}-\frac{\dot{T}}{H}\left(\frac{V'}{V}+\frac{1}{1-\dot{T}^2}\right)\,.\label{tradTeta}
\ea
Equation (\ref{Tepsilon}) shows that tachyonic inflation is similar to $k$-inflation \cite{ADM}. The condition $\epsilon_\H \ll 1$ corresponds to neglect the derivative term in Eq. (\ref{Thj3}) and set $\widetilde{H}^2\approx V^q$: using Eqs. (\ref{FRW}), (\ref{Trho}) and (\ref{Tepsilon}), one gets
\be \label{Thjalt}
V^2=\left(1-\frac{2}{3q}\epsilon_\H\right)\widetilde{H}^{4/q}\,.
\ee
However, the expression for $\eta_\H$ is not very precise from a dynamical point of view because the equation of motion has now a factor $1/(1-\dot{T}^2)$ attached to the second derivative that should be taken into account when neglecting the acceleration term. This suggests to redefine the SR tower by introducing ``covariant derivation'' with respect to the tachyon metric:
\begin{subequations} \label{thsr}
\ba
\bar{\epsilon}_{\text{\tiny $H$},0} &\equiv& \bar{\epsilon}_\H \equiv \epsilon_\H\,,\\
\bar{\epsilon}_{\text{\tiny $H$},n} &\equiv& \prod_{i=1}^n \left\{-\frac{1}{1-\dot{T}^2}\frac{d \ln \left[\left(\frac{\sqrt{1-\dot{T}^2}}{V}\frac{H'}{H^\theta}\right)^{(i-1)}\right]}{d \ln a}\right\}^{1/n}\\
&=& \prod_{i=1}^n \left\{-\frac{1}{1-\frac{2}{3q}\bar{\epsilon}_{\text{\tiny $H$},0}}\frac{d \ln \left[(\sqrt{\bar{\epsilon}_{\text{\tiny $H$},0}})^{(i-1)}\right]}{d \ln a}\right\}^{1/n}\,, \qquad n \geq 1\,,\label{thsrc}
\ea
\end{subequations}
where in the last passage we have used Eqs. (\ref{Tepsilon}) and (\ref{Thj2}).

We can now give a physical interpretation of the tachyon Lagrangian (\ref{ta}), that in our case is
\be \label{Lagr}
{\cal L}= - a^3V\sqrt{1- \epsilon}\,,
\ee
where $\epsilon \equiv 2\epsilon_\H/3q=\dot{T}^2$. The first thing to note is that, if $V\neq 0$, then the Lagrangian is defined only for $\epsilon<1$. This implies that all the cosmologies with $q \leq 2/3$ and a tachyon on the brane with the above non-zero effective Lagrangian experience an accelerated expansion,\footnote{This may be an early-universe inflationary phase as well as the present acceleration period.} while those with $q>2/3$ can be either accelerating or decelerating, depending on the evolution of $\epsilon$. The Gauss-Bonnet high-energy regime is the limiting case; this fact suggested a scenario with an interesting role for the tachyon \cite{LN}, which however seems to have some problems \cite{PS}. Suppose that $q \leq 2/3$ and, at some time $t_*$, the accelerated phase stops, $\epsilon(t_*)=1$: then the tachyon action vanishes. In a string-theoretical framework, when the tachyon reaches the minimum of the potential, the unstable $D$-brane on which it lives annihilates and decays into the closed string vacuum. Put into another language, in the limit $\epsilon\rightarrow 1$, the tachyon metric becomes asymptotically Carrollian, $G_{\mu\nu} \sim -a^2\, \text{diag} (0\,,1\,,1\,,1)$. This property, called Carrollian confinement \cite{GiHY,gib03}, holds for other tachyon effective metrics. Since in the Carroll limit there is no signal propagation, again the string interpretation is that no open tachyonic modes can propagate after the condensation (see also \cite{GuS}).

On the other hand, it can be seen that the vanishing of the Lagrangian (\ref{Lagr}) is not the end of the story by reformulating the theory in the canonical formalism \cite{GHY}. Defining the conjugate comoving momentum density
\be \label{momP}
\Pi_T \equiv \frac{1}{\sqrt{-g}}\frac{\partial {\cal L}}{\partial \dot{T}} = \frac{V\dot{T}}{\sqrt{1-\dot{T}^2}}=\rho\dot{T}\,,
\ee
the density Hamiltonian ${\cal H}=\rho$ in the canonical variables is 
\be
{\cal H} = \Pi_T\dot{T}-{\cal L}= \sqrt{\Pi_T^2 + V^2}\,,
\ee
which is well defined in the condensation limit. Moreover, in string theory the absence of perturbative open modes translates to the fact that, near the minimum of the potential, the string coupling $g_s =O(1)$ and the effective action description might fail down. Possibly, in a cosmological-brane context the vanishing of the tachyon action is a fictitious effect coming from the simplified FRW equation (\ref{FRW}) and the associated dynamics. Actually, a more realistic model would have some implemented mechanism by which, and depending on the position of the minimum of the potential, the consequent cosmological evolution would probably experience a (pre) reheating phase, or a transition to a scalar-driven inflation, in a time interval centered in $t_*$. Similar considerations hold when $q > 2/3$ and the Hubble parameter goes through a boost of the growth rate, recovering late post-inflationary cosmology.

Let us go back to the SR tower. From Eq. (\ref{thsr}) we have
\ba
\bar{\eta}_\H  &\equiv& \bar{\epsilon}_{\text{\tiny $H$},1} = -\frac{1}{1-\dot{T}^2} \frac{\ddot{T}}{H\dot{T}}\,,\label{Tbareta}\\
\bar{\xi}_\H^2 &\equiv& \bar{\epsilon}_{\text{\tiny $H$},2}^2 =  \frac{1}{1-\dot{T}^2}\frac{1}{H^2} \left(\frac{\ddot{T}}{\dot{T}}\right)^. = \frac{1}{(1-\dot{T}^2)^2}\frac{\dddot{T}}{H^2\dot{T}}- \bar{\eta}_\H^2\,.\label{Tbarxi}
\ea
Since $\dot{T} \sim \sqrt{\bar{\epsilon}_\H}$, one can express any $T$-derivative as a time derivative with a purely geometrical factor in front. For example, $\bar{\epsilon}_{\text{\tiny $H$},n}'= \dot{\overline{\epsilon}}_{\text{\tiny $H$},n}\sqrt{3q/2\bar{\epsilon}_\H}$. These expressions carry an extra contribution due to the adopted SR definition, by which $\bar{\eta}_\H\approx O[\bar{\epsilon}_\H(1+\bar{\epsilon}_\H+\cdots)]$. If one wants to keep the spirit of the SR expansion, and neglect by definition these next-to-lowest order terms, one may trade Eqs. (\ref{thsr}) and (\ref{hsr}) for an intermediate definition, by dropping the overall factor in Eq. (\ref{thsrc}),
\begin{subequations} \label{interthsr}
\ba
\epsilon_{\text{\tiny $TH$},0} &\equiv& \epsilon_\H\,,\\
\epsilon_{\text{\tiny $TH$},n} &\equiv& \prod_{i=1}^n \left\{-\frac{d \ln \left[(H'H^{-2})^{(i-1)}\right]}{d \ln a}\right\}^{1/n}\,.
\ea
\end{subequations}
Application of this SR tower will be seen in the following Sections. For the moment, let us note that Eq. (\ref{interthsr}) is formally the same as Eq. (\ref{hsr}) when expressed, with a slight abuse of notation, as a function of the velocity field $\dot{T}$ ($\dot{\phi}$),
\be \label{corresp}
\epsilon_{\text{\tiny $\psi H$},n}^n = \prod_{i=1}^n \frac{-d \ln [\dot{\psi}^{(i-1)}]}{d \ln a}\,.
\ee
Finally, since $\eta_\tH=-\ddot{T}/(H\dot{T})$, one gets
\ba
\dot{\epsilon}_\tH &=& -2 H \epsilon_\tH \eta_\tH\,,\label{Tdotepsi}\\
\dot{\eta}_\tH &=& H\left(\epsilon_\tH\eta_\tH-\xi_\tH^2\right)\,.\label{Tdoteta}
\ea


\subsection{V-SR parameters for a tachyon}

Defining $U\equiv\ln V(T)$, from the SR approximation $\dot{T} \approx -U'/3H$ and Eqs. (\ref{Tepsilon}) and (\ref{Tdotepsi}), we can guess the SR parameters as functions of $V$:
\ba
\epsilon_\tV &\equiv& \frac{q}{6\beta_q^2} \frac{U'^2}{V^q}\,,\\
\eta_\tV &\equiv& -\epsilon_\tV + \frac{1}{3\beta_q^2} \frac{U''}{V^q}\,.
\ea
The complete SR tower comes from the Hubble tower by substituting $\epsilon_\tH$ with $\epsilon_\tV$ and putting $H=\beta_q V^{q/2}$, thus getting
\begin{subequations} \label{Tvsr}
\ba
\epsilon_{\text{\tiny $TV$},0} &\equiv& \epsilon_\tV\,,\\
\epsilon_{\text{\tiny $TV$},n} &\equiv& \frac{1}{3\beta_q^2}\left[\frac{(U')^{n-1}}{V^{nq/2}}\left(\frac{U'}{V^{nq/2}}\right)^{(n)}\right]^{1/n}\,, \qquad n \geq 1\,.
\ea
\end{subequations}
Different SR parameters can be found in \cite{CGJP,SV}. Note that
\ba
\epsilon_\tV' &=& q U' \eta_\tV\,,\\
\eta_\tV' &=& qU'\left(\eta_\tV+\frac{\xi_\tV^2}{2\epsilon_\tV}\right)\,.
\ea


\subsection{SR towers and energy dependence} \label{endep}

It is possible to relate the two SR towers by some simple energy-dependent relations. Here we will restrict ourselves to the first three parameters and define $f=1/3q$. From Eq. (\ref{hjalt}) we get the exact relation
\be \label{VH1}
\epsilon_\sV= \frac{\epsilon_\sH}{9}\,\frac{(3-\eta_\sH)^2}{(1-f\epsilon_\sH)^{1+q}}\,.
\ee
Then, noting that $V'=\dot{\phi}H(\eta_\sH-3)$ and $V''=H^2 [3(\epsilon_\sH+\eta_\sH)-\eta_\sH^2-\xi_\sH^2]$, one has
\be \label{VH2}
\eta_\sV= \frac{(\epsilon_\sH+\eta_\sH)-\frac{1}{3}(\eta_\sH^2+\xi_\sH^2)}{(1-f\epsilon_\sH)^q}\,.
\ee
Finally, noting that $V'''=-3H^3[\theta \epsilon_\sH^2+3\epsilon_\sH\eta_\sH+\xi_\sH^2]+O(\epsilon_\sH^3)$, we obtain, to first H-SR order,
\ba
\epsilon_\sV &\approx& \epsilon_\sH\,,\\
\eta_\sV     &\approx& \eta_\sH+\epsilon_\sH\,,\\
\xi_\sV^2    &\approx& \xi_\sH^2+3\epsilon_\sH\eta_\sH+\theta \epsilon_\sH^2\,.
\ea
These equations allow us to shift from one hierarchy to the other, according to the most convenient approach. Both the SR towers show an explicit dependence on the energy scale because of the definitions, Eqs. (\ref{hsr}) and (\ref{vsr}). Sometimes, this energy dependence can be hidden by proper manipulations of the definitions; however, when differentiating SR parameters, Eqs. (\ref{epsih'}) and (\ref{epsiv'}), the resulting SR combinations contain some factor $q$. Other definitions of the SR tower may have only implicit energy dependence through Eq. (\ref{FRW}); for example, 
\be \label{tow}
\epsilon_n \equiv -\frac{\partial^{n+1} H}{H \partial^{n} H}\,,
\ee
where $\partial^n$ indicates the $n$-th synchronous-time derivative. It turns out that
\be
\dot{\epsilon}_n = H\epsilon_n(\epsilon_0+\epsilon_n-\epsilon_{n+1})\,.
\ee
This tower is related to Eq. (\ref{hsr}), since
\ba
\epsilon_\H &=& \epsilon_0\,,\label{epsi0}\\
2\,\eta_\sH  &=& -\theta\epsilon_0+\epsilon_1\,,\label{epsi1}\\
2\,\xi^2_\sH &=& \epsilon_1\epsilon_2-\epsilon_1^2+\theta\epsilon_0(\epsilon_0-\epsilon_1)\,,
\ea
and so on. Because of the absence of the $1/n$ power, definition (\ref{tow}) does not permit a power truncation similar to that of the traditional SR towers, unless one imposes a constraint such as $\partial^{\bar{n}+1} H=0$ and $\partial^{\bar{n}}H \neq 0$, for some maximum $\bar{n}$.

In the tachyon case, from Eqs. (\ref{Thjalt}), (\ref{Thj2}) and (\ref{Tepsilon}) one has
\be \label{TVH1}
\epsilon_\tV= \frac{\epsilon_\tH}{(1-2f\epsilon_\tH)^{q/2}}\left[1-\frac{\eta_\tH}{6(1-2f\epsilon_\tH)}\right]^2 \,.
\ee
Then, using
\ba
\epsilon_\tH' &=& -H\sqrt{\frac{2\epsilon_\tH}{f}}\eta_\tH\,,\\
\epsilon_\tH'' &=& \frac{H^2}{f} \left(\eta_\tH^2+\xi_\tH^2\right)\,,\\
\eta_\tH' &=& \frac{H}{\sqrt{2f\epsilon_\tH}}\left(\epsilon_\tH\eta_\tH-\xi_\tH^2\right)\,,\\
H' &=& -\sqrt{\frac{\epsilon_\tH}{2f}}H^2\,,
\ea
we get
\be
\eta_\tV= \frac{1}{(1-2f\epsilon_\tH)^{q/2}}\left[\eta_\tH+\frac{2\epsilon_\H\eta_\tH-\eta_\tH^2-\xi_\tH^2}{6(1-2f\epsilon_\tH)}-\frac{(24f+1)\epsilon_\tH\eta_\tH^2}{36(1-2f\epsilon_\tH)^2}\right]\,.
\ee
Hence, to first H-SR order, $\epsilon_\tV \approx \epsilon_\tH$ and $\eta_\tV \approx \eta_\tH$.

In Sec. \ref{vs} we will see one can construct the H-SR tower of the tachyon dynamics from the scalar-field H-SR tower.


\section{\lowercase{e}-foldings and inflationary attractor} \label{attractor}

The number of e-foldings, defined as $N(t)=\int^{t_*}_t H(t') dt'$, measures the amount of inflation from the time $t$, when a comoving perturbation with wavenumber $k(t) =a(t) H(t)$ crosses the horizon, to the end of inflation at $t_*$. A typical ``good'' number of e-foldings is $\approx 50-70$ and many inflationary models have quite a larger total $N$. Sometimes it is useful to perform the integral in the cosmological field; in the scalar field case, from Eq. (\ref{usefull}) one gets
\be \label{N}
N=-\int_{\phi(t)}^{\phi_*}\frac{3q\beta_q^{2-\theta}}{2} \frac{H^{\theta+1}}{H'}\,d\phi= \int_{\phi(t)}^{\phi_*}\left(\frac{3q\beta_q^{2-\theta}}{2} \frac{H^\theta}{\epsilon_\H}\right)^{1/2}\,d\phi\,,
\ee
while for the tachyon case, from Eqs. (\ref{Thj2}) and (\ref{Tepsilon}),
\be \label{TacN}
N=-\int_{T(t)}^{T_*}\frac{3q}{2} \frac{H^3}{H'}\,dT= \int_{T(t)}^{T_*}\left(\frac{3q}{2} \frac{H^2}{\epsilon_\H}\right)^{1/2}\,dT\,.
\ee
Since $k(\psi)=H(\psi)a(\psi)=a_* H(\psi) \exp[N(\psi)]$, the logarithmic scale dependence of the field is exactly
\be
\frac{d \psi}{d\ln k}=\frac{\dot{\psi}}{H(1-\epsilon_\H)}\,,
\label{scdep}
\ee
where $\psi=\phi,\,T$. The predictiveness of inflation depends on the behavior of cosmological solutions with different initial conditions. If there exists an attractor behavior such that the differences of these solutions rapidly vanish, then the inflationary (and post-inflationary) physics will generate observables which are independent of the initial conditions. Let $H_o(\phi)>0$ be a generic expanding solution (denoted with the subscript $o$) of the Hamilton-Jacobi equation (\ref{hj3}) and consider a linear perturbation $\delta H(\phi)$ which does not reverse the sign of $\dot{\phi}>0$. The linearized equation of motion is then
\be
2\sigma_q \widetilde{H}_o' \delta\widetilde{H}'+\widetilde{H}_o^{\theta+1}\delta\widetilde{H}\left[2\theta V\widetilde{H}_o^{\theta-2}-(2+\theta)\right]=0\,,
\ee
where $\sigma_q \equiv 2/(3q\beta_q)^2>0$. Using Eq. (\ref{hjalt}), the solution is
\be
\delta H(\phi) = \delta H(\phi_o)\,\exp \int_{\phi_o}^\phi \frac{d\phi}{2\sigma_q} \left[(2-\theta)\left(1+\frac{\theta}{3}\,\epsilon_\sH\right)\right]\frac{\widetilde{H}_o^{\theta+1}}{\widetilde{H}_o'}\,.\label{attrac}
\ee
The same procedure can be applied to the tachyon equation of motion, giving
\be
\delta H(T) = \delta H(T_o)\,\exp \int_{T_o}^T \frac{dT}{2\sigma_q} \left[(2-\theta)\left(1+\frac{\theta}{3}\,\epsilon_\tH\right)\right]\frac{\widetilde{H}_o^3}{\widetilde{H}_o'}\,.\label{Tattrac}
\ee
All linear perturbations are exponentially damped when the integrand is negative definite and, since $H_o'$ and $\dot{\psi}$ have opposing signs when $q$ is positive, this occurs when the term inside square brackets is positive. There are three cases:
\begin{enumerate}
	\item $0<\theta<2\,\,(q>1)\,:$ It must be $\epsilon_\H>-3/\theta$; this condition is always satisfied, because $\epsilon_\sH$ is positive, and it means that all linear perturbations die away at least exponentially when inflationary solutions approach one another toward the attractor. 
	\item $\theta=0\,\,(q=1)\,:$ The integrand is proportional to $H_o/H_o'<0$ and any linear perturbation is suppressed.
	\item $\theta<0\,\,(0<q<1)\,:$ The damping is achieved when $\epsilon_\H<3/|\theta|$, that is for any inflationary solution with $q>2/5$.
\end{enumerate}
All these cases enclose previous computations in literature: \cite{SB,GPCZ} for the 4D cosmology, \cite{GZZ} for the Randall-Sundrum high-energy regime and \cite{MW} for the full Gauss-Bonnet cosmology. By Eqs. (\ref{N}) and (\ref{TacN}), assuming the slow-roll approximation $\epsilon_\H \approx \text{const}$, the inflationary attractor translates into the condition 
\be \label{damp}
\delta H(\psi) \approx \delta H(\psi_o) \exp \left[-\left(3+\theta\,\epsilon_\H\right) N\right]\,.
\ee
For a given number of e-foldings and $\theta>0$ (RS case), we obtain an enhanced damping with respect to the 4D case, while for $\theta<0$ (GB case) the strength of the attractor is somehow milder. In the case $\theta>2$ ($q<0$), that is when $H_o'$ and $\dot{\psi}$ have concording signs, linear perturbations are suppressed when $\epsilon_\H>-3/\theta$; both the sides of this inequality are negative and in general this relation will not be true. When it is satisfied, we obtain an accelerating universe with both decreasing Hubble length and energy density, that is a super-inflationary universe. For completeness we note that, contrary to what happens in 4D cosmology, for general $q$ it is possible to have both a contracting scale factor and perturbation damping, as it is clear from Eqs. (\ref{attrac}) and (\ref{Tattrac}).

We will not address the issue of how efficient inflation can be; this problem has been studied by many authors under several perspectives. For instance, in the 4D regime, a tachyonic inflationary period turns out to be too short, with a number of e-folding $N =O(10)$ and an early non-linear regime, $\delta\rho/\rho \gg 1$ \cite{CGJP,FKS,KL}. This has suggested the viability of a short tachyonic inflation as a means to provide natural initial conditions for a standard scalar inflationary period,\footnote{Then, this standard inflation lasts a sufficient number of e-folding and dilutes the perturbation structure generated by the tachyonic phase.} similarly to what happens in fast-roll inflation \cite{lin01}. In this sense, a tachyon is not sufficient by itself; nevertheless, the study of its dynamics is worth of investigation, since in other scenarios, such as Randall-Sundrum, things can go better than in the four-dimensional case \cite{BBS,BSS}. It is important to stress again that the analysis of this section is not sufficient to explore all these topics, since we have little constrained the physics involved. This would require the knowledge of the potential and, of course, the gravity framework; perhaps, the most dramatic lack is a condition stating when the confinement of the field on the brane is reliable. These considerations are particularly true in the Gauss-Bonnet scenario, in which the damping condition is critical; see, e.g., \cite{LN,PS}.

As a final remark, we note that Eq. (\ref{damp}) roughly encodes the effects coming from extra dimensions in a term proportional to $\theta$ inside the exponential. For one non-compact extra dimension, this term contributes at most $\pm N$ extra e-foldings, both the sign and magnitude depending on whether the bulk physics in a given energy regime either enhances or opposes the braneworld inflationary expansion. It would be interesting to interpret this result as a general feature of braneworld models and relate the parameter $|\theta|$ to the geometrical setup of the system (number of extra-dimensions and non-compact directions, number of branes and their configuration, etc.); this check would require concrete gravity models with non-standard Friedmann equations, which is beyond the scope of the present work.


\section{Exact cosmological solutions} \label{exact}

So far we have left undetermined the form of the potential $V(\psi)$. Investigation with a few examples shows that, in general, there exists a mapping between scalar and tachyon potentials, in the sense that, chosen a time dependence for the scale factor $a(t)$, from the Hamilton-Jacobi equations (\ref{hjalt}) and (\ref{Thjalt}) there can be found potentials that solves exactly the cosmological equations in the two cases \cite{fei02,pad02,sam03}. We are going to see this in some detail in this section. The scheme to follow is: (1) from $a(t)$, find $H(t)$ and the first SR parameter; (2) from Eqs. (\ref{hjalt}) and (\ref{Thjalt}), find $V(t)$; (3) from Eqs. (\ref{useful1}) and (\ref{Tepsilon}), find $\psi (t)$ and the other SR parameters; (4) substitute $t=t(\psi)$ to find $V(\psi)$. In general, the initial time $t_0$ will not be the origin of time because each solution will be exact in a given patch and not in the entire arc of time from the big bang singularity to, say, the end of inflation. For immediate reference, we summarize the classes of solutions found for the three main energy regimes in Tables \ref{table2}, \ref{table3}, \ref{table4} and \ref{table5}. The space of parameters is chosen in order to have positive $q$, positive potentials,\footnote{Negative potentials have been studied, e.g., in \cite{lin01,FFKL}. Note that the equation of motion of the tachyon possesses a symmetry $V \rightarrow -V$.} real inflaton fields and a strictly expanding universe; contracting cases will be discussed briefly.


\subsection{Scalar field models}

The two classes of models we are going to study have been widely used in literature. Let us start with a power-law scale factor,
\be \label{apl}
a(t)= t^n\,, \qquad H=\frac{n}{t}\,,\qquad n >0\,.
\ee
The SR parameters are
\be \label{plsr}
\epsilon_\sH =q\,\eta_\sH = \sqrt{q}\,\xi_\sH=\frac{1}{n}\,,
\ee
and the potential is
\be
V(t)= \left(1-\frac{1}{3qn}\right)\left(\frac{n}{\beta_q t}\right)^{2/q}\,,\qquad n> \frac{1}{3q}\,.
\ee
Now, since $\dot{\phi} \propto t^{-1/q}$, we must discuss the case $q=1$ separately. If $0<q \neq 1$, then
\be
\phi(t)=\frac{2}{\theta} \left(\frac{2}{3qn}\right)^{1/2}\left(\frac{n}{\beta_q}\right)^{1/q}\,t^{\theta/2}\,,
\ee
and
\be \label{v1}
V(\phi) = A_{q,n}\,\phi^{-4/(q\theta)}\,,
\ee
where $A_{q,n}$ is a coefficient depending on $q$ and $n$. Note that the potential is divergent in $\phi=0$ if $q>1$. 

If $q=1$, we obtain the 4D power-law model \cite{AW,LM2},
\ba
\phi(t) &=& \phi_0+ \left(\frac{2n}{3\beta_1^2}\right)^{1/2} \ln \left(\frac{t}{t_0}\right)\,,\\
V(\phi) &=& \left(1-\frac{1}{3n}\right)\left(\frac{n}{\beta_1 t_0}\right)^2\,\exp\left[-\left(\frac{6\beta_1^2}{n}\right)^{1/2}(\phi-\phi_0)\right]\,.\label{powl}
\ea
There are no contracting solutions.
\begin{table}[ht]
\caption{\label{table2} Exact cosmological solutions for an expanding scale factor $a(t) =t^n$ and a scalar field. Here, $n>1/3q>0$ and $\phi_0=0$. $B$  and $C$ are proportionality coefficients depending on $q$ and $n$.}
\begin{ruledtabular}
\begin{tabular}{ccc}
Regime &     $C\,\phi(t)$    &                   $B\,V(\phi)$              \\ \hline
GB     &      $t^{-1/2}$     &                     $\phi^6$                \\
4D     &      $\ln t/t_0$    &      $\exp (-\sqrt{2\kappa_4^2/n}\,\phi)$   \\
RS2    &       $t^{1/2}$     &                    $\phi^{-2}$              \\
\end{tabular}\end{ruledtabular}
\end{table}

Now, consider a scale factor of the form
\be \label{aexp}
a(t)= \exp (p\, t^n)\,,\qquad H=pn\,t^{n-1}\,,\qquad \text{sgn}(p)=\text{sgn}(n)\,,
\ee
with
\be
V(t)=\left(1+\frac{n-1}{3qpn}\,t^{-n}\right)\left(\frac{pn}{\beta_q}\right)^{2/q}t^{n-2\gamma}\,,
\ee
where $2\gamma=n+2(1-n)/q$; again, $\dot{\phi} =A_{q,p,n} t^{-\gamma}$, with 
\be \label{coeff}
A_{q,p,n} =\left(\frac{2(1-n)\,(pn)^{1-\theta}}{3q\beta_q^{2/q}}\right)^{1/2}\,,
\ee
real if $q>0$ and $n<1$. So,
\be \label{sx}
\epsilon_\sH  =\frac{1-n}{pn}\,t^{-n}\,,\qquad \eta_\sH =\frac{\gamma}{pn}\,t^{-n}\,,\qquad \xi_\sH^2 = \frac{\gamma}{p^2n^2}\,t^{-2n}\,.
\ee
Note that the SR parameters decrease in time and inflation does not naturally end. The reality of the coefficient (\ref{coeff}) guarantees the weak energy condition ($\rho+p \geq 0$, $\rho \geq 0$) if $t_0>\sqrt[n]{(1-n)/(3qpn)}$; from Eq. (\ref{sx}) it then follows that we get inflation from the very beginning only if $q<1/3$. Same considerations are applied for the tachyonic counterpart, with an additional factor $2$ inside the root and a condition $q<2/3$. If $\gamma \neq 1$, then
\ba
\phi(t) &=& \frac{A_{q,p,n}}{1-\gamma}\,t^{1-\gamma}\,,\\
V(\phi) &=& \left[B_{q,p,n}+C_{q,p,n} \phi^{n/(\gamma-1)}\right]\,\phi^{(n-2\gamma)/(1-\gamma)}\,.
\ea
In particular, $q=2$, $\gamma=1/2$ corresponds to the solution for the Randall-Sundrum regime,  while for $q=1$, $\gamma-1=-n/2$, one recovers the 4D intermediate inflation of \cite{bar90,BS}.

If $\gamma =1$, then $0\neq\theta \neq 1$, $n=\bar{n}=\theta/(\theta-1)$ and
\ba
\phi (t) &=& \phi_0 +\bar{A}_{q,p} \ln \left(\frac{t}{t_0}\right)\,,\\
V(\phi) &=& \left\{\bar{B}_{q,p}+\bar{C}_{q,p} \exp \left[-\frac{\bar{n}}{\bar{A}_{q,p}}(\phi-\phi_0)\right]\right\}\exp \left[-\frac{2(1-\bar{n})}{q\bar{A}_{q,p}}(\phi-\phi_0)\right]\,.
\ea
This solution can be applied to just one physically known case, namely, the Gauss-Bonnet regime, with $\bar{n}=1/2$. The contracting solutions are: $p<0$, $0<n\neq1$ (the case $\gamma=1$ is possible only when $0\neq q<1$, $q>2$); $p>0$, $n<0$ (the case $\gamma=1$ is possible only when $1<q<2$). 
\begin{table}[ht]
\caption{\label{table3} Exact cosmological solutions for an expanding scale factor $a(t)=\exp (p t^n)$ and a scalar field, with $\gamma=n/2+(1-n)/q$. Here, $n<1$, $\text{sgn}(p)=\text{sgn}(n)$ and $\phi_0=0$. $B$, $C$ and  $D$ are proportionality coefficients depending on $q$, $n$ and $p$.}
\begin{ruledtabular}
\begin{tabular}{cccc}
Regime &        $\gamma$     &  $C\,\phi(t)$  &                            $B\,V(\phi)$                      \\ \hline
GB     &         $3/2-n$     &  $t^{n-1/2}$   & $\left[1+D\phi^{2n/(1-2n)}\right]\,\phi^{6(n-1)/(2n-1)}$    \\
GB     &           $1$       &  $\ln t/t_0$   & $[1+D\exp(-\frac{C}{2}\phi)]\,\exp(-\frac{3C}{2}\phi)$       \\
4D     &         $1-n/2$     &   $t^{n/2}$    &      $(1+D\phi^{-2})\,\phi^{4(n-1)/n}$               \\
RS2    &          $1/2$      &   $t^{1/2}$    &             $(1+D\phi^{-2n})\,\phi^{2(n-1)}$                 \\
\end{tabular}\end{ruledtabular}
\end{table}


\subsection{Tachyon field models}

With the power law (\ref{apl}), the tachyon field is
\be
T(t) =  \left(\frac{2}{3qn}\right)^{1/2}\,t\,,\qquad \text{sgn}(n)=\text{sgn}(q)\,,
\ee
and the SR parameters read
\be
\epsilon_\tH =\frac{1}{n}\,,\qquad \eta_\tH = \xi_\tH= \cdots=0\,;
\ee
the potential is
\ba 
V &=& \left(1-\frac{2}{3qn}\right)^{1/2}\left(\frac{n}{\beta_q t}\right)^{2/q}\\
  &=& \left(1-\frac{2}{3qn}\right)^{1/2}\left(\frac{2n}{3q\beta_q^2}\right)^{1/q} T^{-2/q}\,,\qquad n> \frac{2}{3q}>0\,.\label{TVpl}
\ea
In order to connect this cosmological solution with string theory, we must take care both of the maximum and the minimum of the potential. As regards the maximum at $T_0=0$, if $q>0$ then the potential (\ref{TVpl}) diverges; as it was shown in \cite{pad02}, it is possible to regularize $V$ and keep an approximated power-law scale factor (\ref{apl}). However, the constance of the kinetic term, $\dot{T} =\sqrt{2/3qn}<1$, which does not satisfy the conditions $\dot{T}(t_0)=0$ and $\dot{T}(t_*)=1$, suggests to regard this solution as an ``intermediate time'' model describing the rolling of the tachyon down its potential, between the very beginning and the asymptotic regime with a pressureless tachyon dust and $n=2/(3q)$. The power-law scalar model with constant SR parameters, Eq. (\ref{plsr}), suffers from the same graceful-exit problem. 
\begin{table}[ht]
\caption{\label{table4} Exact cosmological solutions for an expanding scale factor $a(t) =t^n$ and a tachyon field. Here, $n>2/3q$. $B$  and $C$ are proportionality coefficients depending on $q$ and $n$.}
\begin{ruledtabular}
\begin{tabular}{ccc}
Regime &     $C\,T(t)$    &       $B\,V(T)$           \\ \hline
GB     &      $t$         &       $T^{-3}$            \\
4D     &      $t$         &       $T^{-2}$            \\
RS2    &      $t$         &       $T^{-1}$            \\
\end{tabular}\end{ruledtabular}
\end{table}

In the case of an exponential scale factor, Eq. (\ref{aexp}), the first SR parameters are
\be \label{tx}
\epsilon_\tH  =\frac{1-n}{pn}\,t^{-n}\,,\qquad \eta_\tH =\frac{1}{2p}\,t^{-n}\,,\qquad \xi_\tH^2 = \frac{1}{2p^2n}\,t^{-2n}\,,
\ee
and the potential is
\be
V(t) = \left[1+\frac{2(n-1)}{3qpn}\,t^{-n}\right]^{1/2}\left(\frac{pn}{\beta_q}\right)^{2/q}t^{2(n-1)/q}\,.
\ee
Since $\dot{T}^2 = [2(1-n)/(3qpn)]\,t^{-n}$, one has a real expanding solution when: $p>0$ and $0<n<1$; $p<0$ and $n<0$. If $n \neq 2$, the solution is	
\ba
T(t) &=& \frac{2}{2-n}\left[\frac{2(1-n)}{3qpn}\right]^{1/2} t^{1-n/2}\,,\\
V(T) &=& \left[B_{q,p,n}+C_{q,p,n} T^{2n/(n-2)}\right]^{1/2}\,T^{4(n-1)/[q(2-n)]}\,.
\ea
If $n=2$, then $\dot{T}^2=-t^{-2}/(3qp)$ and $p<0$. The solution is
\ba
T(t) &=& T_0+\left(\frac{-1}{3qp}\right)^{1/2} \ln \left(\frac{t}{t_0}\right)\,,\\
V(T) &=& \left\{1+\frac{1}{3qpt_0^2}\,\exp [-2\sqrt{-3qp}\,(T-T_0)]\right\}^{1/2}\left(\frac{2pt_0}{\beta_q}\right)^{2/q}\exp \left[2\sqrt{\frac{-3p}{q}}\,(T-T_0)\right]\,.
\ea
In the three cosmologies of interest, this solution is contracting. Other exact models can be found in \cite{SV,GKMP}. 
\begin{table}[ht]
\caption{\label{table5} Exact cosmological solutions for an expanding scale factor $a(t)=\exp (p t^n)$ and a tachyon field. Here, $n>2/3q$. $B$, $C$  and $D$ are proportionality coefficients depending on $q$ and $n$.}
\begin{ruledtabular}
\begin{tabular}{ccc}   
Regime &       $C\,T(t)$      &                             $B\,V(T)$                        \\ \hline
GB     &      $t^{1-n/2}$     & $\left[1+DT^{2n/(n-2)}\right]^{1/2}\,T^{6(n-1)/(2-n)}$     \\
4D     &      $t^{1-n/2}$     & $\left[1+DT^{2n/(n-2)}\right]^{1/2}\,T^{4(n-1)/(2-n)}$     \\
RS2    &      $t^{1-n/2}$     & $\left[1+DT^{2n/(n-2)}\right]^{1/2}\,T^{2(n-1)/(2-n)}$     \\
\end{tabular}\end{ruledtabular}
\end{table}

It is possible to relate the solutions of the exponential model (\ref{aexp}) to those of the power-law model. In the former case, the dynamical equations are $V \propto (1+A\,t^{-n})\,t^{2(n-1)/q}$ and $\dot{\psi} \propto t^{-\alpha}$, where $\alpha=-(n/2)+(n-1)/q$ for the scalar field and $\alpha=-n/2$ for the tachyon. In the limit $n \rightarrow 0$, that is when the index of the equation of state $w \rightarrow \text{const}$, both the models formally approach the power-law solution with $V \propto t^{-2/q}$ and $\dot{\phi} \propto t^{-1/q}\,,$ $\dot{T} \propto \text{const}$.


\section{Patch SR formalism and cosmological perturbations} \label{pert}

The advantage of combining the cosmological patch approach with the SR formalism is to provide, at least in certain situations, a unique treatment of physical phenomena for a number of energy regimes. In this section we show an example of this mechanism by discussing the spectra of cosmological perturbations generated by an inflationary era. This topic has been widely (but not completely) explored elsewhere for the scalar field case; a list of references is, e.g., in \cite{mar03}. We use notation and definitions of \cite{lid97}.

Quantum fluctuations of the brane field generate scalar perturbations that can be treated with the linear theory. The main point we will rely on is the independence of the behavior of the curvature perturbations from the gravitational part of the action at sufficiently large scales \cite{WMLL}. It is then possible to consider a linearly perturbed effective 4D metric with the same number of gauge degrees of freedom and borrow part of the standard perturbative formalism. The perturbed metric in the longitudinal gauge reads
\be
d s_4^2\Big|_{brane} \approx a(\tau)^2[(1+2\Phi)~d \tau^2-(1-2\Psi)\delta_{i\!j}~d x^i d x^j]\,,
\ee 
where $\tau=\int dt/a(t)$ is the conformal time and $\Phi$ and $\Psi$ are the gauge-invariant scalar potentials. Assuming that the perturbed effective matter and gravity actions giving the equations of motion have the usual 4D structure, one can perform pure four-dimensional calculations [independent of the FRW equation (\ref{FRW})] and obtain a Mukhanov-type equation. This equation can be solved by an exact cosmological solution with constant slow-roll parameters; from Sec. \ref{exact} we know we have such solutions at our disposal. By perturbing this solution with respect to small variations of the SR parameters, one gets the scalar spectral amplitude. Note that this is not a full ($4+n$)-dimensional derivation, even if its use is well motivated enough. Detailed assumptions and shortcomings of the present procedure in the Randall-Sundrum case can be found, e.g., in \cite{cal1,cal2,RL}.

Given a field $\psi=\phi,\,T$ on the brane, we define the curvature perturbation on comoving hypersurfaces as
\be
{\cal R}=-\Psi-H\frac{\delta\psi}{\dot{\psi}}\,,
\ee
to linear order. The scalar spectrum is 
\be \label{Scsp}
A_s^2 \equiv \frac{4}{25}{\cal P}_\psi \equiv \frac{2k^3}{25\pi^2} \left\langle |{\cal R}_{\mathbf k}|^2\right\rangle\,,
\ee
where angular brackets denote the vacuum expectation value of the perturbation and a bold subscript indicates the $k$-th Fourier mode. 


\subsection{Scalar field case} \label{Apphi}

The Mukhanov equation for the scalar case is \cite{muk85,muk89,MFB}
\be \label{muksc1}
\partial^2_\tau u_{\mathbf k}+\left(k^2-\frac{\partial^2_\tau z}{z}\right)u_{\mathbf k}=0\,,
\ee
where $\partial^2_\tau$ denotes second derivate with respect to conformal time and $u_{\mathbf k}$ are the coefficients of the plane wave expansion of the canonical variable
\be \label{canonical}
u=-z {\cal R}\,.
\ee
Here,
\be
z \equiv \frac{a (\rho+p)^{1/2}}{H}= \frac{a \dot{\phi}}{H}\,.
\ee
Noting that $\partial^2_\tau=a^2(H\partial_t+\partial^2_t)$, it is easy to see that
\ba
\frac{\partial^2_\tau z}{z} &=& a^2H^2\left[(1-\eta_\sH+\epsilon_\sH)(2-\eta_\sH)+\frac{1}{H}(\dot{\epsilon}_\sH-\dot{\eta}_\sH)\right]\nonumber\\
 &=& 2a^2H^2\left[1+\epsilon_\sH-\frac{3}{2}\eta_\sH+O(\epsilon^2)\right]\,,
\ea
where we have isolated the leading order part which is all we need. If the SR parameters are small, then they are constant to leading order because their derivatives are higher order. It is then reasonable to solve the Mukhanov equation with exactly constant SR parameters and perturb the obtained solution. Then, since
\be
\tau = -\frac{1}{aH}\frac{1}{1-\epsilon_\sH}\,,
\ee
Eq. (\ref{muksc1}) can be rewritten as
\be \label{muksc2}
\partial^2_\tau u_{\mathbf k}+\left[k^2-\frac{(\nu^2-1/4)}{\tau^2}\right]u_{\mathbf k}=0\,,
\ee
with
\be
\nu \approx \frac{3}{2}-\eta_\sH+2\epsilon_\sH\,.
\ee
With constant $\nu$, the solution of this equation is $|u_{\mathbf k}| \propto (-\tau)^{1/2} H_\nu^{(1)}(-k\tau)$, where $H_\nu^{(1)}$ is the Hankel function of the first kind of order $\nu$. In the long wavelength limit $k/(aH)\rightarrow 0$, when the mode with comoving wave number $k$ is well outside the horizon, the appropriately normalized solution becomes
\be \label{scasol}
|u_{\mathbf k}| = \frac{2^{\nu-3/2}}{\sqrt{2k}}\frac{\Gamma(\nu)}{\Gamma(3/2)}\left(-k\tau\right)^{-\nu+1/2}\,.
\ee
Expanding the solution (\ref{scasol}) to the same SR order of $\nu$ and substituting it in the previous definition, the scalar spectrum is, to next-to-lowest SR order \cite{SL},
\be
A_s = \left.[1-(2C+1)\epsilon_\sH+C\eta_\sH]\frac{1}{5\pi}\frac{H^2}{\dot{\phi}}\right|_{k=aH}\,;\label{phiS}
\ee
here, $C=\gamma+\ln 2-2 \approx -0.73$ is a numerical constant ($\gamma$ is the Euler-Mascheroni constant) coming from the expansion
\[2^x \frac{\Gamma(x+3/2)}{\Gamma(3/2)} \approx 1-Cx\,,\qquad x \ll 1\,,\] 
and expression (\ref{phiS}) has the asymptotic form at large scales, $k \ll aH$, but is written in terms of quantities evaluated at the horizon crossing of the perturbation. By fixing the term in square brackets equal to 1, one gets the lowest-order expression; this could be directly derived from the fluctuation spectrum of a massless scalar field outside the horizon, by imposing equal-time commutation relations in curved (actually, de Sitter) spacetime \cite{BD},
\begin{subequations} \label{commu}
\ba
\left[\phi ({\bm x}_1,\,t),\,\phi({\bm x}_2,\,t)\right] &=& 0 = \left[\Pi_\phi ({\bm x}_1,\,t),\,\Pi_\phi({\bm x}_2,\,t)\right] \,,\\
\left[\phi ({\bm x}_1,\,t),\,\Pi_\phi({\bm x}_2,\,t)\right] &=& i a^3 \delta^{(3)}({\bm x}_1-{\bm x}_2)\,,
\ea
\end{subequations}
where $\Pi_\phi=\dot{\phi}$ is the conjugate momentum density. The Fourier transform of the fluctuation can be written as a combination of harmonic oscillators, 
\be \label{harmosc}
\delta\phi_{\mathbf k}(t)=w_k(t)a_{\mathbf k}+w_k^*(t)a^\dagger_{-{\mathbf k}}\,,
\ee
where the creation-annihilation operators satisfy the canonical commutation relations, $[a_{{\mathbf k}_1},\,a_{{\mathbf k}_2}^\dagger]=\delta_{{\mathbf k}_1{\mathbf k}_2}$, etc. Near horizon exit, $a = k/H$, and with negligible variation of $H$, the fluctuation amplitude turns out to be $\langle|\delta\phi_{\mathbf k}|^2\rangle = |w_k|^2=H^2/(2k^3)$. Up to numerical factors, $A_s^2 \propto (H^2/\dot{\phi}^2)\,{\cal P}_\phi$, where ${\cal P}_\phi \propto k^3 \langle|\delta\phi_{\mathbf k}|^2\rangle$; therefore, Eq. (\ref{phiS}) is recovered to lowest order. Note that this computation does not involve the Friedmann equation, Eq. (\ref{FRW}), but only the 4D equation of motion (\ref{eom0}).

As a function of the potential, the amplitude (\ref{phiS}) can be written to lowest SR order as \cite{LN}
\be \label{Vampl}
A_s = \frac{3\beta_q^3}{5\pi}\frac{V^{3q/2}}{V'}\,;
\ee
horizon-crossing evaluation is understood. Next-to-lowest order corrections come from the lowest-order relations between H-SR and V-SR towers. The scalar spectral index is defined as
\be
n_s-1 \equiv \frac{d \ln A_s^2}{d \ln k}= \frac{2}{H (1-\epsilon_\H)} \frac{\dot{A}_s}{A_s}\,,
\ee
where we have used Eq. (\ref{scdep}). To first SR-order, the spectral index is uniquely defined for any regime, the energy dependence being relegated to the second order part via Eq. (\ref{epsih'}):
\be \label{n_S}
n_s-1 =\left(2\eta_\sH-4\epsilon_\sH\right)+2(5C+3)\epsilon_\sH\eta_\sH-2C\xi_\sH^2-2[(4-\theta)+2(2-\theta)C]\epsilon_\sH^2\,.
\ee
The running of the scalar index is
\be 
\alpha_s^{(\phi,\theta)} \equiv \frac{d n_s}{d\ln k} \approx 2\left[2(\theta-2)\, \epsilon_\sH^2+5\epsilon_\sH\eta_\sH-\xi_\sH^2\right]\,,\label{alp}
\ee
plus third-order contributions. Equation (\ref{alp}) can be recast as
\be
\alpha_s^{(\phi,\theta)} \approx \epsilon_\sH\left[5(n_s-1)+4(3+\theta)\,\epsilon_\sH\right]\,,\label{alp2}
\ee
assuming that $\xi \ll \text{max}(\epsilon,|\eta|)$. In fact, this approximation can be put into a milder form when considering different patches, $q$ and $q'$, with the same inflaton field, namely, that the parameter $\xi$ is almost constant in the energy regime, $\xi(q) \approx \xi(q')$; this will be sufficient in order to compare theoretical results with observations. In the case of confrontation between a scalar patch and a tachyon patch, it should be $\xi_\phi(q) \approx \xi_T(q')$. We will see in Sec. \ref{vs} that SR parameters of the two scenarios, with the same scale factor, approach one another order by order when $q$ increases; therefore, the last approximation is valid at a certain confidence level if $q,q' \gg 1$. Whichever assumption is chosen to neglect the parameter $\xi$ and write Eq. (\ref{alp}) in terms of observables, it is important to keep in mind that in general $\xi$ cannot be fairly eliminated, even if this is indeed the case in many reasonable situations.


\subsection{Tachyon case} \label{ApT}

In a string-theoretical setup, the quantization of the tachyon Lagrangian (\ref{ta}) is a delicate and non-completely explored subject; in particular, it is not clear yet if the promotion of the classical field to a quantum object correctly describes quantum string theory \cite{sen8}. Nonetheless, one may put aside high-energy motivations for the field theory (\ref{ta}) and study its quantum behavior independently. We start from the commutation relations (\ref{commu}), with $T$ instead of $\phi$ and $\Pi_T$ given by Eq. (\ref{momP}) in the long-wavelength limit. In momentum space, the field fluctuations are quantized as in Eq. (\ref{harmosc}). Now the equation of motion for the perturbation $\delta T_{\mathbf k}$ is not a simple Klein-Gordon equation as in the scalar case; however, one can check that, near horizon crossing, the two-point function of the fluctuation is $\langle|\delta T_{\mathbf k}|^2\rangle = |w_k|^2=H^2/(2Vk^3)$. The computation is performed without adopting any particular gravitation background, except for the hypothesis of quasi\textendash de Sitter expansion. In order to absorb the potential into an expression which is dependent only on the Hubble parameter, we must use Eq. (\ref{FRW}), getting
\be \label{Tspec}
{\cal P}_T \propto k^3 \langle|\delta T_{\mathbf k}|^2\rangle = \frac{\beta_q^{2-\theta}H^\theta}{2c_S}\,,
\ee
where $c_S=\sqrt{-w}=\sqrt{1-\dot{T}^2}$. The presence of an extra $V$-term is also evident when comparing the amplitude $A_s^2 \propto V^{3q+1}/V'^2$ with that in the scalar scenario, Eq. (\ref{Vampl}). Slow-roll corrections to this result can be computed by a more refined treatment including back reaction from the effective 4D metric. The Mukhanov equation for the tachyon case has been derived in a $k$-inflationary context; it reads \cite{MFB,FKS,GM,HN}
\be
\partial^2_\tau u_{\mathbf k}+\left(c_S^2k^2-\frac{\partial^2_\tau z}{z}\right)u_{\mathbf k}=0\,,
\ee
where $u_{\mathbf k}$ are the Fourier coefficients of $u$. In Eq. (\ref{canonical}),
\be
z \equiv \frac{a (\rho+p)^{1/2}}{c_S H}= \frac{1}{\beta_q^{1/q}} \frac{a \dot{T}}{c_S H^{\theta/2}}\,,
\ee
where we have used $\rho+p=\rho \dot{T}^2$. Note that at large scales the curvature perturbation is conserved,
\be
\dot{\cal R} \approx \frac{H}{\rho+p}\,\delta p_\text{nad} =0\,,
\ee
since $\delta\rho/\dot{\rho} =\delta T/\dot{T}$ and the non-adiabatic pressure perturbation $\delta p_\text{nad}\equiv \dot{p}\,[(\delta p/\dot{p})-(\delta\rho/\dot{\rho})]$ vanishes identically. Despite of what happens in the scalar field case, $\theta$ appears explicitly in the definition of $z$. It turns out that
\ba
\frac{\partial^2_\tau z}{z} &=& a^2H^2\left[2+\left(\frac{3\theta}{2}-1\right)\epsilon_\tH-3\eta_\tH+O(\epsilon^2)\right]\,,\\
O(\epsilon^2) &=& \eta_\tH^2 +\xi_\sH^2+ \frac{\theta}{2}\left(\frac{\theta}{2}-1\right)\epsilon_\tH^2-2\theta\epsilon_\tH\eta_\tH+\delta^2[3+(\theta-1)\,\epsilon_\tH-2\eta_\tH+\delta^2]+\frac{(\delta^2)^{^{\bm .}}}{H}\,,
\ea
where $\delta^2 = -\dot{c}_S/(c_SH)=O(\epsilon^2)$. With constant SR parameters, one has
\be \label{mukta2}
\partial^2_\tau u_{\mathbf k}+\left[c_S^2k^2-\frac{(\nu^2-1/4)}{\tau^2}\right]u_{\mathbf k}=0\,,
\ee
with
\be
\nu \approx \frac{3}{2}-\eta_\tH+\left(\frac{\theta}{2}+1\right)\epsilon_\tH\,.
\ee
Again, the solution is of Hankel type but with a rescaled wave number $k \rightarrow c_Sk$. Putting it in the definition of the scalar spectral amplitude (\ref{Scsp}) and tracing $c_S$'s factors yields
\be 
A_s = \left.\left(1-\omega\epsilon_\tH+C\eta_\tH\right)\frac{\beta_q^{1-\theta/2}}{5\pi}\frac{H^{1+\theta/2}}{\dot{T}}\right|_{k=aH}\,,\label{tacS}
\ee
in agreement with Eq. (\ref{Tspec}). Here,
\be
\omega \equiv \left(C+\frac{5}{6}\right)+\frac{\theta}{2}\left(C+\frac{1}{6}\right)\,.
\ee
In the four-dimensional case $\theta=0$, we recover the results of \cite{GM,SV}. To lowest order, it is the same amplitude as that in the non-tachyonic case, if expressed as a function of $H$ and its time derivative:
\be \label{Sdeg}
A_s^2 \approx \frac{3q\beta_q^{2-\theta}}{25\pi^2}\frac{H^{2+\theta}}{2\epsilon_\H}\,.
\ee
This is not surprising, since in the ESR regime the dynamics are almost the same, as explained in Sec. \ref{vs}. The spectral index turns out to be
\be \label{tn_s}
n_s-1 = \left[2\eta_\tH-(2+\theta)\,\epsilon_\tH\right]+2\left(C+1+2\omega\right)\epsilon_\tH\eta_\tH-2C\xi_\tH^2-(2+\theta)\epsilon_\tH^2\,,
\ee
while for the running one gets
\ba
\alpha_s^{(T,\theta)} &\approx& 2\left[(3+\theta)\,\epsilon_\tH\eta_\tH-\xi_\tH^2\right]\label{talp}\\
                      &\approx& (3+\theta)\,\epsilon_\tH\left[(n_s-1)+(2+\theta)\,\epsilon_\tH\right]\,.\label{talp2}
\ea
Note that, to first order, the scalar spectral index is different with respect to the non-tachyon index. In \cite{SV} the lowest-order indices do coincide but because of the use of the horizon-flow parameters $\epsilon_i^{\text{\tiny $(flow)$}}$. As it was shown in \cite{lid03}, these parameters do not properly encode inflationary dynamics even if they provide a good algorithm for reconstructing the inflationary potentials. In standard notation, $\epsilon_1^{\text{\tiny $(flow)$}}=\epsilon_\H$ and either $\epsilon_2^{\text{\tiny $(flow)$}}= (2-\theta)\,\epsilon_\sH-2\eta_\sH$ or $\epsilon_2^{\text{\tiny $(flow)$}}=-2\eta_\tH$, to be used when differentiating Eq. (\ref{Sdeg}). In this sense, our choice of SR parameters better highlights the difference between the two dynamics already to first-order. However, from the point of view of making predictions and comparing them with observations, the H-SR tower and the horizon-flow tower are completely equivalent, as it is shown in the next section.


\subsection{Scalar versus tachyon in 4D and Randall-Sundrum cosmologies: consistency equations} \label{obs}

To compare these models with observations would require the tensor spectrum generated by the gravity sector, which never entered our calculations except for the modified Friedmann equation. In this case it would be possible to construct several ``consistency equations'' relating cosmological observables in a way typical of inflationary scenarios, in which the scalar and gravitational spectra are originated by a unique mechanism. The key point is the knowledge of the behavior of gravity modes on the brane, that is how Kaluza-Klein modes couple with the zero mode confined on the 4D variety. In the standard cosmological scenario, the computation of the tensor spectrum and consistency equations has been performed in a FRW universe to next-to-leading SR order; in the Randall-Sundrum scenario there are only leading-SR-order results \cite{LMW}, while more investigation has been carried on for a marginally perturbed de Sitter brane \cite{KKT,SeT}. Many of the problems arise because of the extra degree of freedom provided by the radion and, in general, a complete solution of the Einstein equations with boundary conditions is difficult to achieve. To leading order, the consistency relation for a scalar-driven inflation is the same in 4D and RS scenarios and a discrimination between them, at least by this method, is not possible. However, quasi-de Sitter computations show a break of the degeneracy and even a possible non-closed structure. Evidences of this in the case of smoothly varying Hubble parameter are also provided \cite{cal2}; it is shown that departures from the standard form of the scalar amplitude would spoil the 4D structure of the consistency relations.

\begin{figure}[ht]
\includegraphics{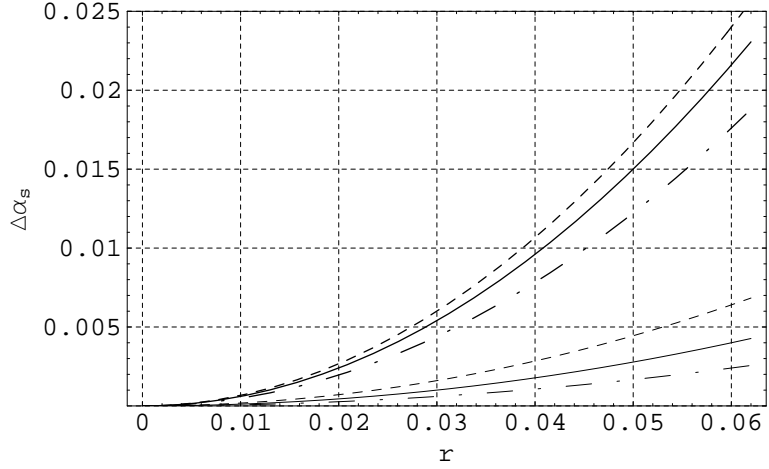}
\caption{\label{fig3} Comparison of the running of the scalar index in several cosmological scenarios with $n_s=1$. Upper group: 4D scalar vs RS tachyon [$\Delta\alpha_s^\text{\tiny $(\phi_0 T_1)$}$, thick dashed line]; 4D scalar vs 4D tachyon [$\Delta\alpha_s^\text{\tiny $(\phi_0 T_0)$}$, thick solid line]; 4D scalar vs RS scalar [$\Delta\alpha_s^\text{\tiny $(\phi_0 \phi_1)$}$, thick dot-dashed line]. Lower group: RS scalar vs RS tachyon [$\Delta\alpha_s^\text{\tiny $(\phi_1 T_1)$}$, thin dashed line]; RS scalar vs 4D tachyon [$\Delta\alpha_s^\text{\tiny $(\phi_1 T_0)$}$, thin solid line]; 4D tachyon vs RS tachyon [$\Delta\alpha_s^\text{\tiny $(T_0 T_1)$}$, thin dot-dashed line].}
\end{figure}

\begin{figure}[ht!]
\includegraphics{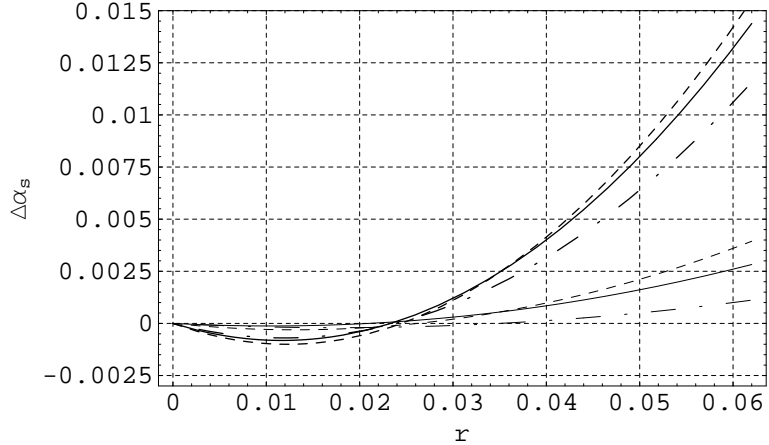}
\caption{\label{fig4} Comparison of the running of the scalar index in several cosmological scenarios with $n_s=0.93$ (best fit of \cite{spe03}): 4D scalar vs RS tachyon [$\Delta\alpha_s^\text{\tiny $(\phi_0 T_1)$}$, thick dashed line]; 4D scalar vs 4D tachyon [$\Delta\alpha_s^\text{\tiny $(\phi_0 T_0)$}$, thick solid line]; 4D scalar vs RS scalar [$\Delta\alpha_s^\text{\tiny $(\phi_0 \phi_1)$}$, thick dot-dashed line]; RS scalar vs RS tachyon [$\Delta\alpha_s^\text{\tiny $(\phi_1 T_1)$}$, thin dashed line]; RS scalar vs 4D tachyon [$\Delta\alpha_s^\text{\tiny $(\phi_1 T_0)$}$, thin solid line]; 4D tachyon vs RS tachyon [$\Delta\alpha_s^\text{\tiny $(T_0 T_1)$}$, thin dot-dashed line].}
\end{figure}

To lowest-order, the squared tensor amplitude in the full Randall-Sundrum model is
\be
A_t^2= \frac{\kappa_4^2}{25\pi^2} \frac{H^2}{2}F^2\left(\frac{H}{\mu}\right)\,, \label{A_T1}
\ee
where the subscript ``$t$'' stands for ``tensor'' and \[F(x)=\left[\sqrt{1+x^2}-x^2 \ln \left(1/x+\sqrt{1+1/x^2}
\right)\right]^{-1/2}\,.\] In the limit $\rho \ll \lambda$, $F(x) \sim 1$ and one recovers the four-dimensional spectrum,
\be
A_{t,4D}^2=\frac{\kappa_4^2}{25\pi^2} \frac{H^2}{2}\,,
\ee
with spectral index
\be
n_t \equiv \frac{ d \ln A_t^2}{d \ln k} \approx -2\epsilon_\H\,;
\ee
in the high-energy limit,
\be \label{Anshel}
A_{t,RS}^2 =  \frac{\kappa_4^2}{25 \pi^2}\frac{3 H^3}{4 \mu}\,,
\ee
where $\mu=\sqrt{\lambda\kappa_4^2/6}\,.$ The spectral index is
\be
n_t  \approx -3\epsilon_\H\,.
\ee
As one can see from Eq. (\ref{Sdeg}), to lowest-order the scalar amplitude does not discriminate between tachyon and non-tachyon inflation; therefore, since the tensor amplitude $A_t$ is independent of the brane content, the lowest-order consistency equation is degenerate with respect to the brane content \cite{SV}. Defining $r=A_t^2/A_s^2$, what comes out is that the consistency equation is
\be
n_t=-2r\,,\qquad \theta=0,1 \qquad \psi=\phi,T\,.
\ee
We do not know if this leading-order degeneracy persists for other braneworld models. The next-to-lowest order consistency relation indeed breaks the degeneracy between scalar field and tachyon field models, but the effect is $O(r^2)$ for $\theta=0$. As explained above, there are no definite results for next-to-leading expressions when $\theta=1$; the procedure of \cite{cal2} brings a deviation from the 4D scenario of order $O(r^2)$ in the case of a scalar field, so the same order of magnitude will characterize the tachyon model. In any case, this effect is too small to be detected by the experiments of this generation, for example the CMB probes WMAP and Planck. A more useful quantity might be the scalar running (the consistency equation for the running of the tensor index is degenerate). In terms of SR parameters, this is a second-order quantity but it comes from the lowest-order part of the scalar amplitude. For the scalar field, we get
\ba
\alpha_s^\text{\tiny $(\phi,0)$} &=& \frac{3r}{3} \left[12r+5 (n_s-1)\right]\,, \label{runs0}\\
\alpha_s^\text{\tiny $(\phi,1)$} &=& \frac{2r}{3}\left[\frac{32}{3}\,r+5 (n_s-1)\right]\,.\label{runs1}
\ea
Let us now use the estimate of observables coming from the first-year data of WMAP \cite{ben03,spe03,bri03}. By inserting the upper bound for the tensor-to-scalar amplitude ratio $r_{max}=0.06$ \cite{ben03} and the best fit for the scalar index of \cite{spe03}, $n_s=0.93$, we find $\alpha_s^\text{\tiny $(\phi,0)$}-\alpha_s^\text{\tiny $(\phi,1)$}=O(10^{-2})$, which is close to the error in the estimate of \cite{bri03}. However, for lower tensor-to-scalar ratios, the effect quits the window of detectability. For the tachyon field,
\ba
\alpha_s^\text{\tiny $(T,0)$}    &=& \frac{9r}{3} \left[2r+ (n_s-1)\right]\,, \label{runt0}\\
\alpha_s^\text{\tiny $(T,1)$}    &=& \frac{8r}{3}\left[2r+ (n_s-1)\right]\,.\label{runt1}
\ea
Unfortunately, $\alpha_s^\text{\tiny $(T,0)$}-\alpha_s^\text{\tiny $(T,1)$}=O(10^{-3})=O(r^2)$ and the tachyon RS scenario does not differ considerably from the four-dimensional one; same order of magnitude for $\alpha_s^\text{\tiny $(\phi,1)$}-\alpha_s^\text{\tiny $(T,i)$}$. On the contrary, a comparison with the 4D scalar field scenario gives $\alpha_s^\text{\tiny $(\phi,0)$}-\alpha_s^\text{\tiny $(T,i)$}=O(10^{-2})$.
We stress once again that the calculations of this section are fully justified only in four dimensions; a complete treatment of cosmological perturbations requires the full Einstein equations with boundary conditions of a given theory of gravity. However, it is instructive to find that scalar-driven and tachyon-driven scenarios lead to different predictions already in the standard 4D model,
\be \label{enfin}
\alpha_s^\text{\tiny $(\phi,0)$} -\alpha_s^\text{\tiny $(T,0)$} = 2r\left[3r+ (n_s-1)\right] \approx 0.01\,.
\ee
With a lower tensor-to-scalar ratio $r$ and a scalar spectrum closer to scale invariance, discrimination between the two scenarios becomes harder to carry on via consistency equations. Define $\Delta\alpha_s^\text{\tiny $(\psi_q \psi'_{q'})$} \equiv \alpha_s^\text{\tiny $(\psi,q)$} -\alpha_s^\text{\tiny $(\psi',q')$}$; for a scale-invariant spectrum ($n_s=1$) and taking the ratio $r = r_{max}/2=0.03$, which is within the $2\sigma$ likelihood bound of \cite{TL}, from Eq. (\ref{enfin}) we have $\Delta\alpha_s^\text{\tiny $(\phi_0 T_0)$} \approx 0.005$, while for $r =  r_{max}/3=0.02$ we get $\Delta\alpha_s^\text{\tiny $(\phi_0 T_0)$} \approx 0.002$, one order of magnitude smaller than the most optimistic high-ratio case. Figures \ref{fig3} and \ref{fig4} show the $\Delta\alpha_s^\text{\tiny $(\psi_q \psi'_{q'})$}$'s for various combinations of models with $n_s=1$ and $n_s=0.93$, respectively.


\section{Relations between scalar-dominated and tachyon-dominated cosmologies} \label{vs}

Because of the non-standard kinetic term in the equation of motion for the tachyon, there is no field redefinition connecting Eqs. (\ref{eom0}) and (\ref{Teom}); in other words, the scalar and tachyon field are dynamically inequivalent. However, it is direct to see that in the extreme slow-roll approximation the two descriptions are not distinguishable to lowest SR order, since, near a local extremum $V\approx \text{const}$, one can rescale $T$ such that $\phi=\sqrt{V} T$ and $V(\phi) \approx V(T)$. This is even clearer looking at the Hamilton-Jacobi equations, Eqs. (\ref{hjalt}) and (\ref{Thjalt}), which are equal up to a second SR order term. For this reason, any cosmological observable generated by an inflationary mechanism with sufficient slow rolling will be rather insensitive to which of the equations of motion is governing the dynamics. As an example, we consider inflationary non-Gaussianity in the Appendix. In general, this first order correspondence between scalar-filled and tachyon-filled backgrounds allows to to relate cosmologies with different index $q$ \cite{muk02}. Thus, one might expect similar predictions for first-order quantities when there is no brane-bulk exchange; however, second order effects may not be irrelevant when comparing the theory with observations, as it has been seen in Sec. \ref{pert}. Outside the SR regime, the tachyon dynamics may lead to qualitatively different scenarios \cite{GKMP}.

Another useful correspondence appears when taking the limit $q \rightarrow \infty$ ($\theta \rightarrow 2$). Then, the SR towers (\ref{hsr}) and (\ref{interthsr}) acquire the same dependence on the Hubble parameter; this fact, together with Eq. (\ref{corresp}), tells us that the inflaton field formally tends to an evolution equation $\dot{\psi} \sim H'/(qH^2)$. Here ``formally'' means that, from a dynamical point of view, this limit is trivial because it forces the inflaton field to a static background [Eqs. (\ref{attrac}) and (\ref{Tattrac}) guarantee that perturbations are frozen]. Nonetheless, if one keeps non-vanishing slow-roll parameters, it can help to derive and check tachyon H-SR tower and formulae from those of the scalar case; see Eqs. (\ref{epsih'}) and (\ref{Tdotepsi}). In fact, general SR combinations will contain $\theta$ factors and remain asymptotically finite; a cross comparison of the slow-roll equations in Secs. \ref{Apphi} and \ref{ApT} nicely shows this feature.

Looking at the exact solutions of the previous section, when going to the limit $q \rightarrow \infty$ in the kinetic term $\dot{\psi}$, scalar solutions approach the tachyon ones within a given background scale factor; in particular, $\alpha \rightarrow -n/2$ and Eq. (\ref{sx}) matches Eq. (\ref{tx}). Using this trick, dynamically inequivalent setups are connected when considering the formal time evolution of the inflaton field with respect to the asymptotic gravitational background. In the horizon-flow language \cite{STG,kin02}, this is equivalent to consider the static solution as the common fixed point of the scalar and tachyon theories, with $\beta$-function given by $\beta \sim \dot{\psi}$ and in the limit in which the horizon-flow tower of the scalar theory approaches the H-SR tower and becomes dynamical.    


\section{Discussion} \label{concl}

In this paper we have explored the behavior of scalar field and tachyon field dynamics by reformulating the Hamilton-Jacobi equations in terms of towers of slow-roll parameters; exact cosmological solutions and the behavior of the inflationary attractor have been provided, as well as the spectra of cosmological perturbations. Scalar-driven and tachyon-driven scenarios are compared and a first-SR-order correspondence is found. In the Appendix some comments on the inflationary bispectrum are made.

It has been assumed that a non-standard Friedmann equation is originated in a non-trivial gravitational context, in which the presence of extra dimensions and the confinement of the visible Universe on a 3-brane determines the effective geometry a brane observer experiences. Indeed this is what explicit gravity models may produce, for example the Randall-Sundrum and the Gauss-Bonnet scenarios. In particular energy regimes, the effective Hubble parameter scales as a power of the energy density on the brane, $\widetilde{H}^2 = \rho^q$; we have restricted our analysis to such a ``patch'' cosmology with constant $q$. This approach does not account for the full structure of the effective Friedmann equation, that must recover the four-dimensional behavior at a sufficiently low density-to-brane tension ratio; however, this is sufficient to formulate most of the non-standard predictions concerning the brane content that arise when the 4D Friedmann equation is violated, $q \neq 1$. On the contrary, quantities such as the gravitational spectrum must be computed with a complete gravity theory at hand. 

With no reference to the gravitational sector, two important assumptions, intimately connected with the evolution of the matter content, emerge; namely, to consider an empty bulk and neglect the Weyl tensor contribution. In particular, there is no source term in the continuity equation (\ref{conti}). In the Randall-Sundrum model, several works have shown that bulk physics mainly affects the small-scale or late-time cosmological structures, i.e., that part of the spectrum which is dominated by post-inflationary physics \cite{GRS,GoM,LCML,IYKOM,koy03}. However, it is possible that a non-zero brane-bulk flux would modify the inflationary spectra. For instance, production of particles when the inflaton does not lie in its vacuum state may generate a non-Gaussianity signature during the accelerated expansion \cite{MRS,GMS}. Cosmic microwave background (CMB) observations strongly constrain the maximum number density of these particles and the $n$-point correlation functions of the resulting perturbations; with a brane-bulk exchange mechanism and interactions at the KK energy scale, this number density, as well as the predicted non-Gaussianity, may vary non-trivially. Thus, the adoption of a modified continuity equation may lead to a richer scenario. See, e.g., \cite{VDMP,LSR,KKTTZ,LMS} for Randall-Sundrum cosmologies with non-diagonal bulk stress-energy tensor and \cite{LT} for a six-dimensional example.

We have focused our attention on patch cosmologies with positive $q$ index, since braneworld models, including the above-mentioned scenarios, generally lay in this range of values. This choice may be justified by other considerations which in turn lead to interesting possibilities. 

First, according to Eq. (\ref{dotH0}), it is not possible to pass from an energy regime A to another B with $\text{sgn}(q_B) =-\text{sgn}(q_A)$ without spoiling the monotonicity of the Hubble parameter. If the dominant energy condition holds, this eventuality might be discarder if one believes in dS-CFT correspondence \cite{str1,SIT,str2,NO,BBM,SSV,kle02,haly1,van04,LVL,haly2,DLS,LuP}, relating time evolution in a de Sitter cosmology to the renormalization group flow of a dual boundary field theory. From a cosmological point of view, the infrared fixed point corresponds to an early inflationary period driven by an effective large cosmological constant $\Lambda_{IR}$ and a small number of effective degrees of freedom, $n_\text{\tiny DOF} \sim 1/\Lambda_{IR}$, while the ultraviolet fixed point is a late-time de Sitter universe with a small cosmological constant $\Lambda_{UV}$ and high number of degrees of freedom. In $dS_4$, the flow is governed by a  monotonically varying $c$-function or central charge $c \propto H^{-2}$. Then, since standard cosmology has $q=1$ and $H>0$, the irreversibility of the flow imposes that $q$ should be always positive. However, apart from intrinsic theoretical problems for this conjectured equivalence, transitions between two such (inflationary) regimes are beyond the scope of the simplified energy-patch cosmology we adopted and, indeed, of the RG approach explored so far (see references for details).\footnote{More generally \cite{bou02}, the holographic principle experiences several difficulties in a cosmological context, in particular with inflation \cite{KaL}.} The author does not know how to reasonably apply such a duality to the present case of an effective 4D quasi--de Sitter brane universe embedded in a higher-dimensional bulk;  therefore, a comparison with a holographic picture appears still unclear and will not be done here.

Secondly, consider the case of a scalar field, Eq. (\ref{rhop}); as discussed in the previous section, an ESR expansion of the energy density yields $\rho^q \propto q \dot{\phi}^2_{eff}/2+V_{eff}$, where the effective theory includes the dimensional contribution of $\beta_q$. If $q<0$, the kinetic term has the wrong sign and this may lead to unitarity problems when quantizing the brane field (also: particles with negative energy propagate forward in time); for a scalar field with positive energy density and equation of state $p=w\rho$, this corresponds to a violation of the null energy condition, $\rho+p \geq 0$, a reasonable assumption according to which light rays are focused by matter. 

However, negative kinetic energies arise in supersymmetric models and higher-derivative-gravity theories \cite{nil84,pol88}, while string models can describe brane physics in which the effective 4D null energy condition is not preserved \cite{CM}; last but not least, anti--de Sitter configurations do violate the dominant energy condition. Until now, the standard lore of well-established energy conditions has been adopted. What about abandoning the old path in favour of more speculative scenarios? In particular, can some of the most popular objections against embarrassing forms of matter be circumvented \cite{pha6}? Recently, many people have been considering scenarios in which the dark energy content of the observable Universe is of a non-conventional nature, namely, violating the null energy condition (see 
\cite{pha1,pha2,pha3,pha4,pha7,pha8,pha11,pha12,pha13,pha14,pha15,pha16,pha17,pha19,pha24,pha25,pha26,pha27} for the case of a non-canonical scalar field and \cite{pha18,pha20,pha23} for the tachyon case). This ``phantom cosmology''\footnote{The zoology of ``phantom physics'' would be quite rich and, indeed, drawn by modifications of just two kinds of fields. Let $[\text{sgn}(\rho),\,\text{sgn}(w+1),\,\text{sgn}(B)]$ describe a field with equation of state $\rho+p=\rho(1+w)$ and kinetic energy proportional to the parameter $B$.  Given a scalar field with energy density $\rho = B\dot{\phi}^2/2+V$, violations of the null energy condition can be achieved with either $[+,-,-]$ (type-I phantoms, commonly known as dark energy ``phantoms'' or ``ghosts'') or $[-,+,-]$ (type-II phantoms). Given a tachyon field with $\rho+p=\rho B\dot{T}^2$, violations corresponds to either $[+,-,-]$ or $[-,+,+]$ (semiphantoms). Conversely, if $\rho+p>0$, then one has either standard matter $[+,+,+]$, type-III phantoms $[-,-,+]$ (scalar case), or superphantoms $[-,-,-]$ (tachyon case).} \cite{pha1}, while successfully exploring previously forbidden regions in the space of parameters, does not avoid criticisms \cite{pha9,pha28}. Anyhow, it would be interesting to ask what is the phenomenology of an early-universe short phantom era with ultra-negative equation of state and generalized Friedmann equation.

From a mathematical point of view, the phantom universe displays interesting properties such as the presence of a finite-time singularity when $w$ in constant [big smash or big rip \cite{pha1,pha5,pha10}, see Eq. (\ref{conw}) with $w<-1$] and a correspondence \cite{chi02,agu03,pha21,pha22} resembling (but not equivalent to) the scale-factor duality of pre-big-bang cosmology which is a symmetry of the low-energy string effective action and is achieved with the mapping $a(t)\rightarrow 1/a(-t)$ \cite{ven91,GV1} (for some reviews on string and pre-big-bang cosmology, see \cite{gas99,LWC,GV2}). Also patch cosmologies with negative $q$ have a finite-time singularity with divergent scale factor, even if the density evolution shows the opposite trend. This fact, together with the non-canonical effective theory which seems to characterize such models, invites to investigate if there is some relation between $q<0$ patch cosmologies and models with phantom fluids or, more generally, if suitable low-energy theories with dilatonic coupling may reproduce this class of effective Friedmann equations.


\begin{acknowledgments}
It is a pleasure to thank Sabino Matarrese and Antonio Riotto for useful comments and suggestions on non-Gaussianity issues. I am also grateful to James Gregory and Antonio Padilla for e-mail discussions on dS and AdS holographies.
\end{acknowledgments}


\appendix*

\section{Slow-roll correspondence and inflationary non-Gaussianity} \label{gauss}

According to the inflationary paradigm, small quantum fluctuations of the inflaton field are amplified to cosmological scale by the accelerated expansion. These perturbations then leave their imprint into the cosmic microwave background as thermal anisotropies. Two main physical observables generated by this mechanism are the scalar spectrum, which is (the Fourier transform of) the two-point correlation function of scalar perturbations, and the bispectrum, coming from the three-point function. For the gravitational potential $\Phi$, this reads
\be
\langle\Phi({\mathbf k}_1)\Phi({\mathbf k}_2)\Phi({\mathbf k}_3)\rangle = (2\pi)^3\delta^3\left({\mathbf k}_1+{\mathbf k}_2+{\mathbf k}_3\right) \sum_{i<j}f_{NL}({\mathbf k}_i, {\mathbf k}_j) {\cal P}_\Phi(k_i){\cal{P}}_\Phi(k_j)\,,
\ee
where ${\cal P}_\Phi$ is the gravitational spectrum and sum indices run from 1 to 3. The non-linearity parameter $f_{NL}$ allows to write the gravitational potential in terms of its linear Gaussian part $\Phi_L$; in real space,
\be
\Phi({\bm x})=\Phi_L({\bm x}) +\frac{1}{2}f_{NL}\left[\Phi^2_{L}({\bm x})- \left\langle \Phi^2_{L}({\bm x})\right\rangle \right]\,.
\ee
If the statistical distribution is Gaussian, $f_{NL}=0$, then the three-point function vanishes.

In four dimensions, it turns out that the inflationary contribution to the non-linear parameter is $f_{NL}^\phi=O(n_s-1) =O(\epsilon)$ for a single scalar field \cite{ABMR,mal03}. The above considerations suggest that a similar effect should arise in the tachyon-driven scenario. We have verified this by using the stochastic approach of Gangui \textit{et al.} \cite{GLMM}. This approach permits us to estimate the order of magnitude of the effect by just considering second-order fluctuations of a self-interacting inflaton field and no gravitational fluctuations. This might seem too crude an approximation, since one should go up to second order in perturbation theory in order to fully take gravitational back-reaction into account and treat the bispectrum consistently. Surprisingly, the inflaton perturbation really encodes the main feature of the model apart from the resulting incorrect combination of SR parameters. So, $f_{NL}^T = O(\epsilon)$. On the other side, even if a full second-order calculation for tachyon fluctuations would provide the exact result, one should note that the post-inflationary era greatly enhances non-Gaussianity, up to $f_{NL}^{\text{\tiny post}} =O(1)$ \cite{BMR1,BMR2}. Therefore, the effect is subdominant.

Actually, we have performed the calculation with the general FRW equation (\ref{FRW}) and the V-SR tower both for the scalar field and the tachyon. However, in the presence of extra dimensions the gravitational contribution may lead to a non-trivial behavior of second order perturbations, since to this order the interplay between extra-horizon scales and small scales may become quite delicate. This would impose a more rigorous treatment and a full second-order calculation in order to carefully evaluate non-Gaussianity produced during inflation.


\end{document}